\documentclass[]{iopart}
\usepackage{graphicx}
\usepackage{iopams}

\begin{document}

\title[Spectral analysis for cascade-emission-based quantum communication]
{Spectral analysis for cascade-emission-based quantum communication in atomic ensembles}

\author{H. H. Jen}
\address{Physics Department, National Tsing Hua University, Hsinchu 300, Taiwan, R. O. C.}
\ead{sappyjen@gmail.com}

\begin{abstract}
The ladder configuration of atomic levels provides a source for telecom photons (signal) from the upper atomic transition. \ For rubidium and cesium
atoms, the signal field has the range around 1.3-1.5 $\mu$m that can be coupled to an optical fiber and transmitted to a remote location. \ Cascade
emission may result in pairs of photons, the signal entangled with the subsequently emitted infrared photon (idler) from the lower atomic transition.\  This correlated two-photon source is potentially useful in the (Duan-Lukin-Cirac-Zoller) DLCZ protocol for the quantum repeater.\  We implement the cascade emission to construct a modified DLCZ quantum repeater and investigate the role of time-frequency entanglement in the protocol.\ The dependence of protocol on photon-number resolving and non-resolving detectors is also studied.\  We find that frequency entanglement deteriorates the performance but the harmful effect can be diminished by using shorter pump pulses to generate the cascade emission.\ An optimal cascade-emission-based DLCZ scheme is realized by applying a pure two-photon source in addition to using detectors of perfect quantum efficiency.
\end{abstract}
\pacs{42.50.Ex, 42.50.Dv}
\submitto{\jpb}
\maketitle
\section{Introduction}
Quantum communication has opened up the possibility to transmit quantum information over long distance.  A quantum repeater protocol proposed by Briegel {\it et al.} \cite{repeater, Dur} fulfills such a long distance system.  Subsequently, Duan, Lukin, Cirac, and Zoller (DLCZ) \cite{dlcz} suggested a long distance quantum communication based on atomic ensembles. \ This scheme involves Raman scattering of an incoming light from the atoms with the emission of a signal photon. \ The photon is then correlated with coherent excitation of the atomic ensemble.  The information may be transfered through light to another atomic ensemble or retrieved by a reverse Raman scattering process, generating an idler photon directional correlated with the signal one \cite{qubit, chou, store, pan, kimble}.  The signal and idler photons in alkali gases are in the near-infrared spectral region, which mismatches the telecommunication bandwidth optical fiber.  Therefore, an alternative process that is able to generate telecom wavelength photons correlated with atomic spin excitations \cite{telecom, radaev, conversion} would provide the essential step toward practical long distance quantum communication.

The alkali atomic cascade transition shown in figure \ref{four} is able to generate telecom wavelength light, the signal, from the upper transition and a near-infrared field, the idler, from the lower one.  The telecom light can travel through the fiber with minimal loss, while the near-infrared field is suitable for storage and retrieval in an atomic quantum memory element. \ Their use in a quantum information system requires quantum correlations between stored excitations and the telecom field. \ It is interesting to assess the cascade scheme in the DLCZ protocol given that it could potentially reduce transmission losses in a quantum telecommunication system. \ 

Correlated photon pairs may be generated by parametric down conversion (PDC) \cite{clock,pulsepump,branning}. \ The degree of entanglement can be
quantified by Schmidt mode decomposition \cite{law,parker}, allowing the influence of group-velocity matching \cite{eliminate} to be assessed. \ A pure
single photon source is a basis element for quantum computation by linear optics (LOQC) \cite{LOQC}, and it can be conditionally generated by
measurement \cite{single}. \ A similar approach can be applied to the study of the transverse degrees of freedom in type-II PDC \cite{transverse} and PDC in
a distributed microcavity \cite{micro}. \ In photonic-crystal fiber (PCF), a factorizable photon pair can be generated by spectral engineering
\cite{fiber}. The spectral effect has been discussed in relation to a quantum teleportation protocol \cite{spectral} as a first step toward quantum communication.

This motivates the research in this article where we study the spectral effect of correlated photon pair generated from cascade atomic ensemble in DLCZ scheme.  
The DLCZ scheme is based on entanglement generation and swapping, and quantum state transfer, which make up the basic elements for long distance quantum communication.  Generating entanglement is the first step in quantum information processing, and entanglement swapping is the essence to distribute the entanglement over distant places.  Quantum state transfer enables the secure transmission to eavesdropping which is therefore of great practical interest \cite{QI, cryp, QI2}.

In this article, we start from formulating a two-photon state generated in a cold atomic ensemble.\  We use Schr\"{o}dinger's equation to investigate the correlated signal and idler photons spontaneously emitted from two driving lasers via four-level atomic structure.\  The essence of phase-matching in the four-wave mixing (FWM) conditions is discussed in section 2, and we calculate the second-order correlation function to show the bunching behavior of the cascade-emitted photons.  \ In section 3, we briefly review the Schmidt decomposition that is used to analyze the frequency entanglement and mode functions of the two-photon source.\  We characterize these spectral properties of the correlated two-photon state for different superradiant decay constants and study how the laser excitation pulse modifies their spectral profile.\  We then demonstrate the modified DLCZ scheme, a quantum repeater protocol, which employs cascade emission in section 4.\  We reconstruct the elements of DLCZ scheme including entanglement generation, entanglement swapping, effective `polarization' maximally entangled state projection, and quantum teleportation. \ We investigate how frequency entanglement of the cascade photon pair influences these elements, and study their performances (fidelity, heralding and success probabilities) for two types of photon detectors (resolving photon number or not) along with dependence of quantum efficiency.\  We then conclude in section 5 and discuss the alternative method of generating telecom photon by frequency conversion. \ The details of the Hamiltonian and Schr\"{o}dinger's equation are described in appendix A.\  In appendix B, we derive the multimode two-photon state and conditional output density operators used in modified DLCZ protocol. 

\section{A correlated two-photon state}

\begin{figure}[t]
\begin{center}
\includegraphics[
natheight=7.499600in,
natwidth=9.999800in,
height=2.5037in,
width=3.5682in]{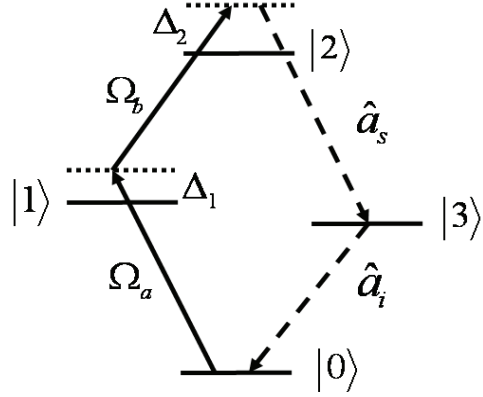}
\caption{Four-level atomic ensemble interacting with two driving lasers
(solid) with Rabi frequencies $\Omega_{a}$ and $\Omega_{b}.$ \ Cascade emissions, signal and
idler fields, are labelled by $\hat{a}_{s}$ and $\hat{a}_{i},$ respectively and
$\Delta_{1}$ and $\Delta_{2}$ are single and two-photon laser detunings. }%
\label{four}%
\end{center}
\end{figure}

We consider $N$ cold atoms that are initially prepared in the ground state interacting with four independent electromagnetic fields.\ \ As shown in
figure \ref{four}, two driving lasers (of Rabi frequencies $\Omega_{a}$ and $\Omega_{b}$) excite a ladder configuration $|0\rangle\rightarrow
|1\rangle\rightarrow|2\rangle.$ \ Two quantum fields, signal $\hat{a}_{s}$ and idler $\hat{a}_{i},$ are generated spontaneously.\ The four atomic levels can be chosen as ($|0\rangle$, $|1\rangle$, $|2\rangle$, $|3\rangle$)=($|5\textrm{S}_{1/2},\textrm{F}=3\rangle$, $|5\textrm{P}_{3/2},\textrm{F}=4\rangle$, $|4\textrm{D}_{5/2},\textrm{F}=5\rangle$, $|5\textrm{P}_{3/2},\textrm{F}=4\rangle$) \cite{telecom}.\ The atoms adiabatically follow the two excitation pulses and decay through the cascade emission of signal and idler photons.\ \ Based on the discussion in appendix A, we permit only single atomic excitations under the condition of large detuning, $\Delta_{1}\gg\sqrt{N}\Omega_{a}/2$. \ The Hamiltonian and the coupled equations of the atomic dynamics are detailed in appendix A. \ %

To correctly describe the frequency shifts arising from dipole-dipole interactions, we need to include the non-rotating wave approximation (non-RWA) terms in the electric dipole interaction Hamiltonian. \ In appendix A, we consider only RWA terms for simplicity and the non-RWA terms would allow virtual transitions that in effect add to the frequency shifts in an appropriate way \cite{Lehm,Scully09}.  The frequency shift has contributions from the single atom Lamb shift and a collective frequency shift. \ The Lamb shift is assumed to be renormalized into the single atom transition frequency distinguishing it from the collective shift due to the atom-atom interaction.

Writing the state-vector $|\psi(t)\rangle$ in a basis restricted to single atomic excitations, and single pairs of signal and idler photons, we can
introduce the probability amplitudes,%

\begin{eqnarray}
C_{s,k_{i}}(t)=\sum_{\mu=1}^{N}e^{-i\vec{k}_{i}\cdot\vec{r}_{\mu}}%
\langle3_{\mu},1_{k_{s},\lambda_{s}}|\psi(t)\rangle,
\end{eqnarray}
and%

\begin{eqnarray}
D_{s,i}(t)=\langle0,1_{k_{s},\lambda_{s}},1_{k_{i},\lambda_{i}}|\psi(t)\rangle,
\end{eqnarray}
where $\vec{k}_i$ here is equivalent to $\vec{q}_i$ defined in appendix A and is denoted as $\vec{k}_a+\vec{k}_b-\vec{k}_s$ from FWM condition which will be demonstrated later. \ Note that $C_{s,k_{i}}(t)$ is an amplitude for a phased excitation of the ensemble of atoms subsequent to signal photon emission and ($k_{s,i}$, $\lambda_{s,i}$) represent wave vectors and polarization indices for signal and idler fields respectively.

After adiabatically eliminating the laser excited levels in the equations of motion, we are able to simplify and derive the amplitude $C_{s,k_{i}}$\ and
the signal-idler (two-photon) state amplitude $D_{s,i}$ as shown in appendix A,%

\begin{eqnarray}\fl
C_{s,k_{i}}(t)=g_{s}^{\ast}(\epsilon_{k_{s},\lambda_{s}}^{\ast}\cdot\hat
{d}_{s})\sum_{\mu}e^{i\Delta\vec{k}\cdot\vec{r}_{\mu}}\int_{0}^{t}dt^{\prime
}e^{i(\omega_{s}-\omega_{23}-\Delta_{2})t^{\prime}}e^{(-\frac{\Gamma_{3}^{N}%
}{2}+i\delta\omega_{i})(t-t^{\prime})}b(t^{\prime}),
\end{eqnarray}

\begin{eqnarray}\fl
D_{s,i}(t) =g_{i}^{\ast}g_{s}^{\ast}(\epsilon_{k_{i},\lambda_{i}}^{\ast
}\cdot\hat{d}_{i})(\epsilon_{k_{s},\lambda_{s}}^{\ast}\cdot\hat{d}_{s}%
)\sum_{\mu}e^{i\Delta\vec{k}\cdot\vec{r}_{\mu}}\int_{0}^{t}\int_{0}%
^{t^{\prime}}dt^{\prime\prime}dt^{\prime}e^{(-\frac{\Gamma_{3}^{N}}{2}%
+i\delta\omega_{i})(t^{\prime}-t^{\prime\prime})}\nonumber\\
 e^{i(\omega_{i}-\omega_{3})t^{\prime}}e^{i(\omega_{s}-\omega_{23}-\Delta
_{2})t^{\prime\prime}}b(t^{\prime\prime}),
\end{eqnarray}
where $b(t)=\frac{\Omega_{a}(t)\Omega_{b}(t)}{4\Delta_{1}\Delta_{2}}$ is proportional to the product of the Rabi frequencies.\ Coupling constants $g_{s(i)}$, polarization direction $\epsilon_{k_{s(i)},\lambda_{s(i)}}$, and unit direction of dipole operators $\hat{d}_{s,i}$ are for signal and idler fields respectively.\ Various definition of optical frequencies $\omega$'s and laser detuning $\Delta_2$ can be found in Appendix A.

The factor $\sum_{\mu}e^{i\Delta\vec{k}\cdot\vec{r}_{\mu}}$ reflects phase-matching of the interaction under conditions of four-wave mixing when
the wavevector mismatch $\Delta\vec{k}=\vec{k}_{a}+\vec{k}_{b}-\vec{k}%
_{s}-\vec{k}_{i}\rightarrow0.$ \ The radiative coupling between atoms results in the appearance of the superradiant decay constant%

\begin{eqnarray}
\Gamma_{3}^{N}=(N\bar{\mu}+1)\Gamma_{3},%
\end{eqnarray}
where $\Gamma_{3}$ is the natural decay rate of the $|3\rangle\rightarrow |0\rangle$ transition, and $\bar{\mu}$ is a geometrical constant depending on
the shape of the atomic ensemble. \ An expression for the collective frequency shift $\delta\omega_{i}$\ is given in the Appendix A.  For a cylindrical atomic ensemble, the decay factor $N\bar{\mu}+1$ depends on the height and radius as shown in equation (\ref{mu}). \ $N\bar{\mu}+1\approx4$ and $6$ which are comparable to the operating conditions of the experiment \cite{telecom}.

We use normalized Gaussian pulses as an
example where $\Omega_{a}(t)=\frac{1}{\sqrt{\pi}\tau}\tilde{\Omega}_{a}e^{-t^{2}/\tau^{2}}$,$~\Omega_{b}(t)=\frac{1}{\sqrt{\pi}\tau}\tilde
{\Omega}_{b}e^{-t^{2}/\tau^{2}}$, so that the two pulses are overlapped with the same pulse width. $\ \tilde{\Omega}_{a,b}$ is the pulse area, and let
$\Delta\omega_{s}\equiv\omega_{s}-\omega_{23}-\Delta_{2}-\delta\omega_{i},~\Delta\omega_{i}\equiv\omega_{i}-\omega_{3}+\delta\omega_{i}$. \ In the long time limit, we have the probability amplitude $D_{si}$,%

\begin{equation}\fl
D_{si}(\Delta\omega_{s},\Delta\omega_{i})=\frac{\tilde{\Omega}_{a}%
\tilde{\Omega}_{b}g_{i}^{\ast}g_{s}^{\ast}(\epsilon_{k_{i},\lambda_{i}}^{\ast
}\cdot\hat{d}_{i})(\epsilon_{_{s}}^{\ast}\cdot\hat{d}_{s})}{4\Delta_{1}%
\Delta_{2}}\frac{\sum_{\mu}e^{i\Delta\vec{k}\cdot\vec{r}_{\mu}}}{\sqrt{2\pi
}\tau}\frac{e^{-(\Delta\omega_{s}+\Delta\omega_{i})^{2}\tau^{2}/8}}%
{\frac{\Gamma_{3}^{N}}{2}-i\Delta\omega_{i}},\label{longtwo}%
\end{equation}
indicating a spectral width $\Gamma_{3}^{N}/2$ for idler photon in a Lorentzian distribution modulating a Gaussian profile with a spectral width
$2\sqrt{2}/\tau$ for signal and idler. \ Energy conservation of signal and idler photons with driving fields at their central frequencies corresponds to
$\omega_{s}+\omega_{i}=\omega_{a}+\omega_{b}$, which makes $\Delta\omega_{s}+\Delta\omega_{i}=0$; the collective frequency shifts cancel.

Using the asymptotic form of the two-photon state given in equation (\ref{longtwo}), the second-order correlation function \cite{QO:Scully} is calculated as
\begin{eqnarray}
G_{s,i}^{(2)}=\langle\psi(\infty)|\hat{E}_{s}^{-}(\vec{r}_{1},t_{1})\hat{E}_{i}^{-}(\vec{r}_{2},t_{2})\hat{E}_{i}^{+}(\vec{r}_{2}%
,t_{2})\hat{E}_{s}^{+}(\vec{r}_{1},t_{1})|\psi(\infty)\rangle
\end{eqnarray}
where

\begin{eqnarray}
\hat{E}_{s}^{+}(\vec{r}_{1},t_{1})&=\sum_{k_{s},\lambda_s}\sqrt{\frac
{\hbar\omega_{s}}{2\epsilon_{0}V}}\hat{a}_{k_{s},\lambda_s}\vec{\epsilon}%
_{k_{s},\lambda_{s}}e^{i\vec{k}_{s}\cdot\vec{r}_{1}-i\omega_{s}t_{1}},\\
  \hat{E}_{i}^{+}(\vec{r}_{2},t_{2})&=\sum_{k_{i},\lambda_i}\sqrt{\frac
{\hbar\omega_{i}}{2\epsilon_{0}V}}\hat{a}_{k_{i},\lambda_i}\vec{\epsilon}%
_{k_{i},\lambda_{i}}e^{i\vec{k}_{i}\cdot\vec{r}_{2}-i\omega_{i}t_{2}}.%
\end{eqnarray}

$|\psi(\infty)\rangle$ denotes the state vector in the long time limit that involves the ground state and two-photon state vectors. \ Free
electromagnetic fields, signal and idler photons, at space ($\vec{r}_{1},\vec{r}_{2}$) and time ($t_{1},t_{2}$) are $\hat{E}_{s}^{+}$ and $\hat{E}%
_{i}^{+}$ where ($+$) denotes their positive frequency part. \ For second order correlation function, only $D_{si}$ contributes to it, then following the standard procedure of $G_{s,i}^{(2)}$ calculation \cite{QO:Scully}, we have%

\begin{eqnarray}
\sqrt{G_{s,i}^{(2)}}\propto\sum_{\mu}e^{i\Delta\vec{k}\cdot\vec{r}_{\mu}}e^{-2(\Delta t_{s})^{2}/\tau^{2}}e^{-\frac{\Gamma_{3}^{N}}{2}(\Delta
t_{i}-\Delta t_{s})}\Theta(\Delta t_{i}-\Delta t_{s}),\label{g2}%
\end{eqnarray}
where $\Delta t_{s}\equiv t_{1}-\frac{\vec{r}_{1}\cdot\hat{k}_{s}}{c}$ and $\Delta t_{i}\equiv t_{2}-\frac{\vec{r}_{2}\cdot\hat{k}_{i}}{c}$.  The step function $\Theta$ shows the causal connection between signal and idler emissions and is due to the complex integral with the pole at $\Delta\omega_{i}=-i\frac
{\Gamma_{3}^{N}}{2}-\delta\omega_{i}$ in the lower half plane. \ The emission time for the signal field ($t_{1}-\frac{\vec{r}_{1}\cdot\hat{k}_{s}}{c}$) is within the pulse envelope of width $\tau$, and the idler photon decays with a superradiant constant $\Gamma_{3}^{N}/2$.

If we let $\Delta t\equiv\Delta t_{i}-\Delta t_{s}$ and choose $\Delta t_{s}=0$ as the origin in time (idler gating time), then we have the
second-order correlation function%

\begin{equation}
G_{s,i}^{(2)}(\Delta t)=|\Phi_{s,i}(\Delta t)|^{2}\propto e^{-\Gamma_{3}%
^{N}\Delta t}\textrm{ where }\Delta t\geq0.
\end{equation}
\ It resembles the result for the second-order correlation function in the case of single atom, whereas here we have an enhanced decay rate due to the
atomic dipole-dipole interaction.  This exponential correlation function indicates the bunching property of cascade photons \cite{QL:Loudon} showing an immediate emission of idler photon following the signal one.

\section{Schmidt decomposition}

We would like to perform an analysis of entanglement properties of our cascade emission source. \ In addition to polarization entanglement, a
characterization of frequency space entanglement is required to clarify its suitability in the DLCZ protocol \cite{dlcz}.

In the long time limit, the state function is given by%

\begin{equation}
|\psi\rangle=|0,\textrm{vac}\rangle+\sum_{s,i}D_{s,i}|0,1_{\vec{k}_{s},\lambda_{s}},1_{\vec{k}_{i},\lambda_{i}}\rangle
\end{equation}
where $D_{s,i}$ can be found in equation (\ref{longtwo}) and $|0,$vac$\rangle$ is the joint atomic ground and photon vacuum state.\ Shorthand notations $s=(k_{s},\lambda_{s})$ and $i=(k_{i},\lambda_{i})$ are for different spatial modes $k_{s(i)}$ and two degree of freedom polarizations $\lambda_{s(i)}$.

The spatial correlation of two-photon state in FWM condition can be eliminated by pinholes or by coupling to single mode fiber so we consider only the continuous frequency space. \ For some specific polarizations $\lambda_{s}$ and $\lambda_{i}$, we have the state vector $|\Psi\rangle$,%

\begin{equation}
|\Psi\rangle=\int f(\omega_{s},\omega_{i})\hat{a}_{\lambda_{s}}^{\dag}%
(\omega_{s})\hat{a}_{\lambda_{i}}^{\dag}(\omega_{i})|0\rangle d\omega
_{s}d\omega_{i},
\end{equation}
where
\begin{equation}
f(\omega_{s},\omega_{i})=\frac{e^{-(\Delta\omega_{s}+\Delta\omega_{i})^{2}%
\tau^{2}/8}}{\frac{\Gamma_{3}^{N}}{2}-i\Delta\omega_{i}}.\label{longtwo2}%
\end{equation}

Following the theoretical work on two-photon pulses generated from down-conversion by Law {\it et al.} \cite{law}, the quantification of entanglement can be determined in the Schmidt basis where the state vector is expressed as%

\begin{eqnarray}
& |\Psi\rangle=\sum_{n}\sqrt{\lambda_{n}}\hat{b}_{n}^{\dag}\hat{c}_{n}^{\dag
}|0\rangle,\\
& \hat{b}_{n}^{\dag}\equiv\int\psi_{n}(\omega_{s})\hat{a}_{\lambda_{s}}^{\dag
}(\omega_{s})d\omega_{s},\\
& \hat{c}_{n}^{\dag}\equiv\int\phi_{n}(\omega_{i})\hat{a}_{\lambda_{i}}^{\dag
}(\omega_{i})d\omega_{i},
\end{eqnarray}
where $\hat{b}_{n}^{\dag},$ $\hat{c}_{n}^{\dag}$ are effective creation operators and $\lambda_n$'s (no confusion with polarization index $\lambda_{s,i}$) are probabilities in corresponding two-photon emission modes $n$.\ If $\lambda_1=1$, it means a pure two-photon emission.\ Eigenvalues $\lambda_{n}$, and eigenfunctions $\psi_{n}$ and $\phi_{n},$ are the solutions of the eigenvalue equations,%

\begin{eqnarray}
\int K_{1}(\omega,\omega^{\prime})\psi_{n}(\omega^{\prime})d\omega^{\prime}  &
=\lambda_{n}\psi_{n}(\omega),\\
\int K_{2}(\omega,\omega^{\prime})\phi_{n}(\omega^{\prime})d\omega^{\prime}  &
=\lambda_{n}\phi_{n}(\omega),
\end{eqnarray}
where $K_{1}(\omega,\omega^{\prime})\equiv\int f(\omega,\omega_{1})f^{\ast
}(\omega^{\prime},\omega_{1})d\omega_{1}$ and $K_{2}(\omega,\omega^{\prime
})\equiv\int f(\omega_{2},\omega)f^{\ast}(\omega_{2},\omega^{\prime}%
)d\omega_{2}$ are the kernels for the one-photon spectral correlations
\cite{law,parker}. \ Orthogonality of eigenfunctions is $\int\psi_{i}%
(\omega)\psi_{j}(\omega)d\omega=\delta_{ij}$, $\int\phi_{i}(\omega)\phi
_{j}(\omega)d\omega=\delta_{ij},$ and the normalization of quantum state requires $\sum_{n}\lambda_{n}=1$.

In the Schmidt basis, the von Neumann entropy may be written%

\begin{equation}
S=-\sum_{n=1}^{\infty}\lambda_{n}\textrm{ln}\lambda_{n}.
\end{equation}

If there is only one non-zero Schmidt number $\lambda_{1}=1$, the entropy is zero, which means no entanglement and a factorizable state. \ For more than
one non-zero Schmidt number, the entropy is larger than zero and bipartite entanglement is present.

The kernel in equation (\ref{longtwo2}) has all the frequency entanglement information, entanglement means $f(\omega_{s},\omega_{i})$ cannot be
factorized in the form $g(\omega_{s})h(\omega_{i}),$ a multiplication of two separate spectral functions. \ By inspection the Gaussian profile of signal
and idler emission is a source of correlation. \ The joint spectrum $\Delta\omega_{s}+\Delta\omega_{i}$ is confined within the width of order of
$1/\tau$. \ The Lorentzian factor associated with the idler emission has a width governed by the superradiant decay rate.%

\begin{figure}[t]
\begin{center}
\includegraphics[
natheight=6.000400in,
natwidth=7.099700in,
height=2.425in,
width=3.4601in]%
{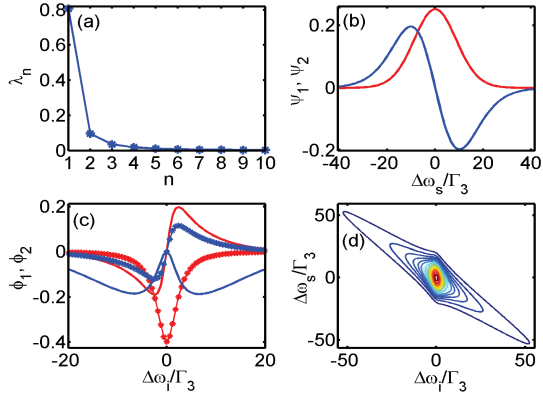}%
\caption{Schmidt mode analysis with pulse width $\tau=0.25$ and superradiance
decay factor $N\bar{\mu}+1=5.$ \ (a) Schmidt number and (b) signal mode
functions: Re[$\psi_{1}$] (solid-red) and Re[$\psi_{2}$] (solid-blue).
Imaginary parts are not shown, then are zero. (c) Real (solid) and imaginary
(dotted) parts of first (red) and second (blue) idler mode functions,
$\phi_{1}$ and $\phi_{2}$. \ (d) The absolute spectrum $|f(\Delta\omega
_{s},\Delta\omega_{i})|$.}%
\label{spec_mode}%
\end{center}
\end{figure}

In figure \ref{spec_mode}, we show the Schmidt decomposition of the spectrum.
\ We use a moderate superradiant decay constant $N\bar{\mu}+1=5,$ comparable
to the reference \cite{telecom}, and a nanosecond pulse duration
$\tau=0.25~(\textrm{in units of }1/\Gamma\approx 26 \textrm{ ns}$), and $\Gamma_{3}/2\pi=6$ MHz. \ Due to
slow convergence associated with the Lorentzian profile, we use a frequency
range up to $\pm1200$ (in unit of $\Gamma_{3}$) with $2000\times2000$ grid.
\ The numerical error in the eigenvalue calculation is estimated to be about
$1\%$ error. \ In this case, the largest Schmidt number is $0.8$ and
corresponding signal mode function has a FWHM Gaussian profile $4\sqrt{2\textrm{
ln}(2)}/\tau\approx 19 \Gamma_{3}$. \ The idler mode function $\phi_{1}%
$\ reflects the Lorentzian profile in the spectrum at the signal peak
frequency ($\Delta\omega_{s}=0$),%

\begin{equation}
f(\Delta\omega_{s}=0,\Delta\omega_{i})=\frac{e^{-\Delta\omega_{i}^{2}\tau
^{2}/8}}{(N\mu+1)\Gamma_{3}/2-i\Delta\omega_{i}}%
\end{equation}
where a relatively broad Gaussian distribution is overlapped with a narrow
spread of superradiant decay rate [FWHM $>(N\bar{\mu}+1)\Gamma_{3}/2$].

\begin{figure}[t]
\begin{center}
\includegraphics[
natheight=6.000400in,
natwidth=7.099700in,
height=2.5745in,
width=3.5036in
]%
{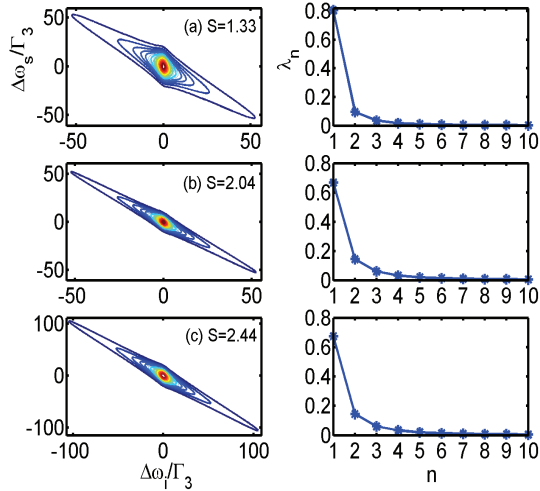}%
\caption{Absolute spectrum of two-photon state and the eigenvalues of Schmidt
decomposition. $N\bar{\mu}+1=5$ for both (a) $\tau=0.25$ (b) $\tau=0.5$.
$N\bar{\mu}+1=10$ for (c) $\tau=0.25$. \ The von Neumann entropy (S) is
indicated in the plots.}%
\label{spec_schmidt}%
\end{center}
\end{figure}

Figure \ref{spec_schmidt} shows that the cascade emission source is more
entangled if the superradiant decay constant, or the pulse duration increases.
\ We note that the Gaussian profile aligns the spectrum along the axis
$\Delta\omega_{s}=-\Delta\omega_{i}$ and the spectral width for signal photon
at the center of the idler frequency distribution ($\Delta\omega_{i}=0$) is
determined by pulse duration $\tau$. \ For a shorter pulse $\tau^{-1}%
>(N\bar{\mu}+1)\Gamma_{3}/2$, the joint Gaussian profile has a larger width,
and the spectrum is cut off by the Lorentzian idler distribution. \ A larger
width leads to a less entangled source and distributes the spectral weight
mainly along the crossed axes $\Delta\omega_{s}=0$ and $\Delta\omega_{i}=0$.
\ A narrow Lorentzian profile cuts off the entanglement source term
$e^{-(\Delta\omega_{s}+\Delta\omega_{i})^{2}\tau^{2}/8}$ tilting the spectrum
along the line $\Delta\omega_{s}+\Delta\omega_{i}=0.$ \ In the opposite limit,
$\tau^{-1}<(N\bar{\mu}+1)\Gamma_{3}/2$, the spectrum is highly entangled
corresponding to tight alignment along the axis $\Delta\omega_{s}%
=-\Delta\omega_{i}$ (figure \ref{spec_schmidt} (c)).%

Note that the short pulse duration ($\tau\geq0.25$ ($6.5$ ns)) should not violate the assumption of adiabaticity $\tau\gtrsim1/\Delta_{1}\textrm{ or
}1/\Delta_{2}$.

The Schmidt analysis and calculation of von Neumann entropy shows that signal-idler fields are more entangled if the ensemble is more optically
dense, corresponding to stronger superradiance. \ For the DLCZ protocol, we wish to avoid frequency entanglement. The superradiance may be reduced with
smaller atomic densities but good qubit storage and retrieval efficiency require a moderate optical thickness \cite{telecom}. \ A better approach
involves using short pulse excitation $\tau^{-1}>(N\bar{\mu}+1)\Gamma_{3}$. We will investigate the spectral properties in more details for the DLCZ scheme
in the next Section.\ Note that there has been a development in the setting of spontaneous parametric downconversion to generate frequency-uncorrelated entangled photons by using shorter pump pulses for scalable all-optical quantum information processing \cite{Pan_SPDC}.

\section{DLCZ scheme with cascade emission}

In the DLCZ protocol, a weak pump laser Raman scatters a single photon generating a quantum correlated spin excitation in the ensemble. \ By
interfering the Raman photons generated from two separate atomic ensembles on a beam splitter (B.S.), the DLCZ entangled state $(|01\rangle+|10\rangle
)/\sqrt{2}$ \cite{0110} is prepared conditioned on one and only one click of the detectors after the B.S. \ Hence $|0\rangle$ and $|1\rangle$ represent the
state of zero or one collective spin excitations stored in the hyperfine ground state coherences. \ This state originates from indistinguishable
photon paths. \ The error from multiple excitations can be made negligible if the pump laser is weak enough. \ 

\begin{figure}[t]
\begin{center}
%\ifcase\msipdfoutput
\includegraphics[
natheight=5.599600in,
natwidth=7.099800in,
height=2.5208in,
width=3.5192in]%
{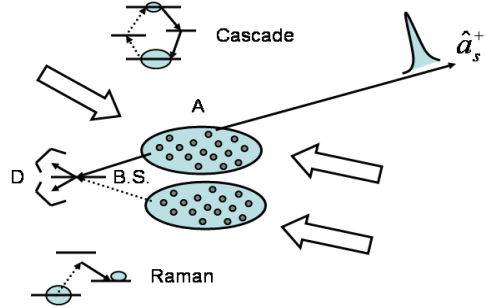}%
\caption{Entanglement generation in the DLCZ scheme using the cascade and
Raman transitions in two different atomic ensembles. \ Large white arrows
represent laser pump excitations corresponding to the dashed lines in either
cascade or Raman level structures. \ Here $\hat{a}_{s}^{\dagger}$ represents
the emitted telecom photon. \ B.S. means beam splitter that is used to
interfere the incoming photons measured by the photon detector D. \ The label
A refers to the pair of ensembles for later reference.}%
\label{ent_gen}%
\end{center}
\end{figure}

As shown in figure \ref{ent_gen}, we consider instead that one of the ensembles employs cascade emission. \ The idea is for cascade emission to
generate a telecom photon ($\hat{a}_{s}^{\dag}$) for transmission in the optical fiber, and an infrared photon that interferes locally with the Raman
photon generated in the $\Lambda$-type atomic ensemble. \ In this way interference of the infrared photons generate the entangled state,%

\begin{equation}
|\Psi\rangle=\frac{1}{\sqrt{2}}(|01\rangle_{a,s}+|10\rangle_{a,s}),
\end{equation}
similar to the conventional DLCZ entanglement generation scheme. \ Now, however, instead of a stored spin excitation, we generate a telecom photon. \ Here we denote $|\Psi\rangle$ as a matter-light entangled state where (a, s) represent an atomic collective spin excitation and a telecom photon respectively.  

The entanglement swapping with the cascade emission may be implemented as shown in figure \ref{ent_swap}, and will be discussed in detail in the next
Section. \ The initial state is a tensor product of two state vectors generated locally at the sites A and B.%
\begin{eqnarray}\fl
|\Psi\rangle_{AB}=(\sqrt{1-\eta_{1A}}|0\rangle+\sqrt{\eta_{1A}}|1\rangle
_{i}^{A}|1\rangle_{s}^{A})\otimes(\sqrt{1-\eta_{2A}}|0\rangle+\sqrt{\eta_{2A}%
}|1\rangle_{r}^{A}|1\rangle_{a}^{A})\otimes\nonumber\\\fl
 (\sqrt{1-\eta_{1B}}|0\rangle+\sqrt{\eta_{1B}}|1\rangle_{i}^{B}|1\rangle
_{s}^{B})\otimes(\sqrt{1-\eta_{2B}}|0\rangle+\sqrt{\eta_{2B}}|1\rangle_{r}%
^{B}|1\rangle_{a}^{B}),\label{joint}%
\end{eqnarray}
where (s,~i) represent the signal and idler photons from the cascade emission, and (r, a) are Raman scattered photon and the collective spin excitation.
\ Here$\ \eta_{1}$ and $\eta_{2}$ are efficiencies to generate cascade and Raman emission. \ Since $\eta_{1}$ and $\eta_{2}\ll1,$ multiple atomic
excitations or multi-photon generation can be excluded.

\subsection{Entanglement swapping}

\begin{figure}[t]
\begin{center}
%\ifcase\msipdfoutput
\includegraphics[
natheight=5.499600in,
natwidth=7.099800in,
height=2.5436in,
width=3.5856in]%
{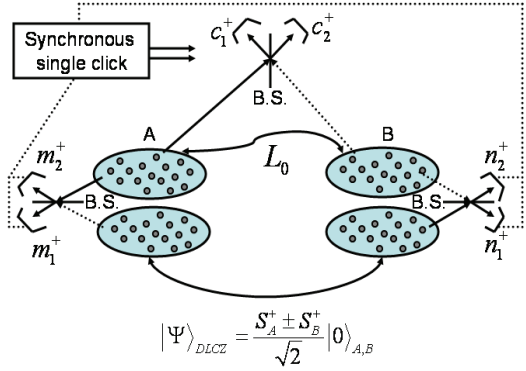}%
\caption{Entanglement swapping of DLCZ scheme using the cascade transition.
The site A is described in detail in figure \ref{ent_gen} and equivalently for
the site B. \ The telecom signal photons are sent from both sites and
interfere by B.S. midway between with detectors represented by $c_{1}%
^{\dagger}$ and $c_{2}^{\dagger}$. \ Synchronous single clicks of the
detectors from both sites ($m_{1,2}^{\dagger}$, $n_{1,2}^{\dagger}$) and the
midway detector ($c_{1,2}^{\dagger}$) generate the entangled state between
lower atomic ensembles at sites A and B. \ The locally generated entanglement
is swapped to distantly separated sites in this cascade-emission-based DLCZ
protocol.}%
\label{ent_swap}%
\end{center}
\end{figure}

Before we proceed to expand the product state of equation (\ref{joint}) and investigate the spectral effects of cascade emission on the modified DLCZ scheme, we would like to address the intrinsic errors from the protocols.\ Consider the product state generated from entangled states of A and B as in figure \ref{ent_swap},%

\begin{eqnarray}\fl
|\Psi\rangle_A\otimes|\Psi\rangle_B  &=(\frac{|10\rangle_{as}+|01\rangle_{as}}{\sqrt{2}})_{A}%
\otimes(\frac{|10\rangle_{as}+|01\rangle_{as}}{\sqrt{2}})_{B}\nonumber\\\fl
 &=\frac{1}{2}(|1010\rangle_{asas}+|1001\rangle_{asas}+|0110\rangle
_{asas}+|0101\rangle_{asas}),
\end{eqnarray}
where the subscript (a) represents a stored local atomic excitation, and (s)
means a telecom photon propagating toward the B.S. in the middle. \ We can
tell from this effective state that the first component ($|1010\rangle_{asas}%
$) contributes no telecom photons at all (two local excitations) and can be
ruled out by measuring a "click" at one of the middle detectors. \ The second
and the third components have components of the entangled state of quantum
swapping, and the fourth one is the source of error if the photodetector
cannot resolve one from two photons. \ The error could be corrected by using a
photon number resolving detector (PNRD) if other drawbacks like dark counts,
photon losses during propagation, and detector inefficiency are not considered.

Now we will formulate the entanglement swapping including the spectral effects
discussed in Section 3. \ We ignore pump-phase offsets, assuming $50/50$ B.S.
and a symmetric set-up ($\eta_{1A}=\eta_{1B}=\eta_{1},~\eta_{2A}=\eta
_{2B}=\eta_{2}$) for simplicity. \ Expand the previous joint state, equation (\ref{joint}) and keep the terms up to the second order of $\eta_{1,2}$ that
can contribute to detection events ($\hat{m}_{1,2}^{\dag},\hat{n}_{1,2}^{\dag
}$),%

\begin{eqnarray}\fl
|\Psi\rangle_{eff}&  =\eta_{1}(1-\eta_{2})|1\rangle_{i}^{A}|1\rangle_{s}^{A}|1\rangle_{i}%
^{B}|1\rangle_{s}^{B}+\eta_{2}(1-\eta_{1})|1\rangle_{r}^{A}|1\rangle_{cs}%
^{A}|1\rangle_{r}^{B}|1\rangle_{cs}^{B}\nonumber\\\fl
&+  \sqrt{\eta_{1}\eta_{2}(1-\eta_{1})(1-\eta_{2})}|1\rangle_{i}^{A}%
|1\rangle_{s}^{A}|1\rangle_{r}^{B}|1\rangle_{cs}^{B}\nonumber\\\fl &+\sqrt{\eta_{1}\eta
_{2}(1-\eta_{1})(1-\eta_{2})}|1\rangle_{r}^{A}|1\rangle_{cs}^{A}|1\rangle
_{i}^{B}|1\rangle_{s}^{B},\label{mode}%
\end{eqnarray}
where the cascade emission state $|1\rangle_{s}|1\rangle_{i}\equiv\int
f(\omega_{s},\omega_{i})\hat{a}_{\lambda_{s}}^{\dag}(\omega_{s})\hat
{a}_{\lambda_{i}}^{\dag}(\omega_{i})|0\rangle d\omega_{s}d\omega_{i}$ has the
spectral distribution $f(\omega_{s},\omega_{i})$\ as derived in Section 3.

As shown in figure \ref{ent_swap}, entanglement swapping protocol is fulfilled
by measuring three clicks from the three pairs of the detectors respectively
($\hat{m}_{1,2}^{\dag},\hat{n}_{1,2}^{\dag},\hat{c}_{1,2}^{\dag}$). \ The
quantum efficiency of the detector is considered in the protocol, and we
describe a model for quantum efficiency in Appendix B.1. \ We then use this
model to describe photodetection events registered by non-resolving photon
detectors (NRPD). \ Starting with the input density operator $\hat{\rho}%
_{in}=|\Psi\rangle_{eff}\langle\Psi|,$ we derive the projected density
operator, equation (\ref{out}), conditioned on the three clicks of $\hat{m}%
_{1}^{\dag},\hat{n}_{1}^{\dag},$ and $\hat{c}_{1}^{\dag}$ in Appendix C.2. We
use the Schmidt decomposition of the projected density operator and assume a
single mode for the Raman scattered photon. \ We find the un-normalized
density operator $\hat{\rho}_{out}^{(2)}$\ given in equation (\ref{out}),%

\begin{eqnarray}\fl
\hat{\rho}_{out}^{(2)}=&  \frac{\eta_{1}^{2}(1-\eta_{2})^{2}}{16}(2-\eta_{t})\eta_{t}\eta_{eff}%
^{2}\Big(1+\sum_{j}\lambda_{j}^{2}\Big)|0\rangle\langle0|+\frac{\eta_{1}%
\eta_{2}(1-\eta_{1})(1-\eta_{2})}{8}\eta_{t}\eta_{eff}^{2}\nonumber\\\fl
&  \bigg\{\Big(\hat{S}_{B}^{\dag}|0\rangle\langle0|\hat{S}_{B}+\hat{S}%
_{A}^{\dag}|0\rangle\langle0|\hat{S}_{A}\Big)+\sum_{j}\lambda_{j}\int\phi
_{j}(\omega_{i})\phi_{j}^{\ast}(\omega_{i}^{\prime})\Phi^{\ast}(\omega
_{i})\Phi^{\ast}(\omega_{i}^{\prime})d\omega_{i}d\omega_{i}^{\prime
}\nonumber\\\fl
&  \Big(\hat{S}_{B}^{\dag}|0\rangle\langle0|\hat{S}_{A}+\hat{S}_{A}^{\dag
}|0\rangle\langle0|\hat{S}_{B}\Big)\bigg\},\label{out3}%
\end{eqnarray}
where $\eta_{t}$ and $\eta_{eff}$ are quantum efficiencies of the detectors at the telecom and infrared wavelengths respectively.\ $\lambda_j$'s are Schmidt eigenvalues derived in section 3.\ The first term in equation (\ref{out3}) is the atomic vacuum state at sites A and B and contributes an
error to the output density operator. \ The second term contains the
components of the DLCZ entangled state.

We can define the fidelity $F$ as the projection of density operator to the entangled state $|\Psi\rangle_{DLCZ}=(S_{A}^{\dag}+S_{B}^{\dag})|0\rangle/\sqrt{2}$ and calculate the success probability $P_{S}$ of entanglement swapping of the entangled state and the heralding probability $P_{H}$ for the third click as \cite{shapiro}

\begin{eqnarray}
F  \equiv\frac{\Tr(\hat{\rho}_{out}^{(2)}|\Psi\rangle_{DLCZ}%
\langle\Psi|)}{\Tr(\hat{\rho}_{out}^{(2)})},\\
P_{H}   =P_{1}+P_{2},~P_{1}=P_{2}=\frac{\Tr(\hat{\rho}_{out}^{(2)}%
)}{\mathcal{N}},\\
P_{S}   =P_{1}\times F_{1}+P_{2}\times F_{2},~F_{1}=F_{2}=F,
\end{eqnarray}
where $P_{1,2}$ is the heralding probability of the single click from the
midway detector ($\hat{c}_{1,2}^{\dag})$ as shown in figure \ref{ent_swap},
and a trace (Tr) is taken over atomic degrees of freedom. \ The normalization
factor $\mathcal{N}$\ is calculated in equation (\ref{normalization}) and is given
by
\begin{equation}\fl
\mathcal{N}=\frac{\eta_{1}^{2}(1-\eta_{2})^{2}}{4}\eta_{eff}^{2}+\frac
{\eta_{1}\eta_{2}(1-\eta_{1})(1-\eta_{2})}{2}\eta_{eff}^{2}+\frac{\eta_{2}%
^{2}(1-\eta_{1})^{2}}{4}\eta_{eff}^{2}.
\end{equation}

We have used the following properties for the calculation of $\hat{\rho}%
_{out}^{(2)}$ and $\mathcal{N}$,%

\begin{equation}
\int d\omega_{s}d\omega_{i}|f(\omega_{s},\omega_{i})|^{2}=1,
\end{equation}
where orthonormal relations in the mode functions are used, and%
\begin{equation}
\int d\omega_{s}d\omega_{s}^{\prime}d\omega_{i}d\omega_{i}^{\prime}%
f(\omega_{s}^{\prime},\omega_{i}^{\prime})f^{\ast}(\omega_{s}^{\prime}%
,\omega_{i})f(\omega_{s},\omega_{i})f^{\ast}(\omega_{s},\omega_{i}^{\prime
})=\sum_{j}\lambda_{j}^{2}.
\end{equation}
Note that the single mode spectral function for the Raman photon satisfies
$\int d\omega|\Phi(\omega)|^{2}=1.$

The fidelity, heralding, and success probability become
\begin{eqnarray}
 F=\frac{1+\sum_{j}\lambda_{j}\int\phi_{j}(\omega_{i})\phi_{j}^{\ast}%
(\omega_{i}^{\prime})\Phi^{\ast}(\omega_{i})\Phi^{\ast}(\omega_{i}^{\prime
})d\omega_{i}d\omega_{i}^{\prime}}{\eta_{r}(2-\eta_{t})(1+\sum_{j}\lambda_{j}^{2})/2+2},\\
 P_{H}=\frac{\eta_{r}\eta_{t}(2-\eta_{t})(1+\sum_{j}\lambda_{j}^{2}%
)/2+2\eta_{t}}{(\sqrt{\eta_{r}}+1/\sqrt{\eta_{r}})^{2}},\\
 P_{S}=\eta_{t}\frac{1+\sum_{j}\lambda_{j}\int\phi_{j}(\omega_{i})\phi
_{j}^{\ast}(\omega_{i}^{\prime})\Phi^{\ast}(\omega_{i})\Phi^{\ast}(\omega
_{i}^{\prime})d\omega_{i}d\omega_{i}^{\prime}}{(\sqrt{\eta_{r}}+1/\sqrt
{\eta_{r}})^{2}},
\end{eqnarray}
where $\frac{1-\eta_{2}}{1-\eta_{1}}\approx1$ and $\eta_{r}=\eta_{1}/\eta_{2}$.

The fidelity depends on a sum of square of Schmidt numbers in the denominator
and the mode mismatch between the idler and Raman photons in the numerator.
\ Let us assume that the Raman photon mode is engineered to be matched with
the idler photon mode of the largest Schmidt number ($\phi_{1}(\omega_{i})$ in
our case), which is required to have a larger fidelity (so is the success
probability) compared to other modes. \ We may also compare the NRPD with the
performance of PNRD in the midway detectors, then we have the fidelity,
heralding, and success probability,%

\begin{eqnarray}
F   =\left\{
\begin{array}
[c]{c}%
\frac{1+\lambda_{1}}{\eta_{r}(2-\eta_{t})(1+\sum_{j}\lambda_{j}^{2}%
)/2+2},~\textrm{NRPD}\\
\frac{1+\lambda_{1}}{\eta_{r}(1-\eta_{t})(1+\sum_{j}\lambda_{j}^{2}%
)+2},~\textrm{PRND}%
\end{array}
\right. \label{fidelity}\\
P_{H}   =\left\{
\begin{array}
[c]{c}%
\frac{\eta_{r}\eta_{t}(2-\eta_{t})(1+\sum_{j}\lambda_{j}^{2})/2+2\eta_{t}%
}{(\sqrt{\eta_{r}}+1/\sqrt{\eta_{r}})^{2}},~\textrm{NRPD}\\
\frac{\eta_{r}\eta_{t}(1-\eta_{t})(1+\sum_{j}\lambda_{j}^{2})+2\eta_{t}%
}{(\sqrt{\eta_{r}}+1/\sqrt{\eta_{r}})^{2}},~\textrm{PRND}%
\end{array}
\right. \label{herald}\\
P_{S}   =\left\{
\begin{array}
[c]{c}%
\frac{\eta_{t}(1+\lambda_{1})}{(\sqrt{\eta_{r}}+1/\sqrt{\eta_{r}})^{2}%
},~\textrm{NRPD}\\
\frac{\eta_{t}(1+\lambda_{1})}{(\sqrt{\eta_{r}}+1/\sqrt{\eta_{r}})^{2}%
},~\textrm{PRND}%
\end{array}
\right.  .\label{success}%
\end{eqnarray}

When the relative efficiency is made arbitrarily small, the fidelity
approaches $(1+\lambda_{1})/2$ for both types of detectors. \ It reaches one
if a pure cascade emission source is generated (von Neumann entropy $E=0$ and
$\lambda_{1}=1$). \ When $\eta_{r}=1$ with a pure source using NRPD with a
perfect quantum efficiency, $F=2/3,~P_{H}=3/4,~P_{S}=1/2,$ which coincide with
the results of the reference \cite{shapiro} (with perfect quantum efficiency).

We discuss the frequency entanglement for various pulse widths and superradiant decay rates in Section 3. \ We find that for shorter driving
pulses and smaller superradiant decay rates, the cascade emission source is less spectrally entangled. \ That means when $\eta_{r}$ is fixed, a shorter
driving pulse heralds a higher fidelity DLCZ entangled state.%

\begin{figure}[t]
\begin{center}
\includegraphics[
natheight=6.000400in,
natwidth=7.099700in,
height=2.5386in,
width=3.5636in]%
{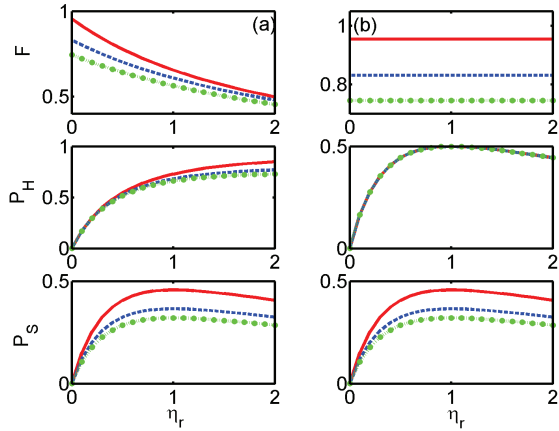}%
\caption{Fidelity $F$, heralding $P_{H}$, and success $P_{S}$ probabilities of
entanglement swapping versus relative efficiency $\eta_{r}$\ with perfect
detection efficiency $\eta_{t}=1.$ \ Column (a) NRPD and (b) PNRD.
\ Solid-red, dashed-blue, and dotted-green curves correspond to the pulse
width parameters $\tau=(0.1,0.5,0.5)$ and superradiant factor $N\bar{\mu
}+1=(5,5,10)$ (see Section 3 and Appendix A)$.$ \ The von Neumann entropy is
$S=(0.684,2.041,2.886),$ respectively.}%
\label{F_etar}%
\end{center}
\end{figure}

In figure \ref{F_etar}, we numerically calculate the entropy and plot out the fidelity from equation (\ref{fidelity}), the heralding probability from equation
(\ref{herald}), and the success probability from equation ( \ref{success}) as a function of the relative efficiency $\eta_{r}.$ \ With a perfect detection
efficiency ($\eta_t=1$), we find that at a smaller $\eta_{r},$ the less entangled source gives us a higher fidelity DLCZ entangled state but with a
smaller success probability. \ Small generation probability for cascade emission ($\eta_{r}<1$) reduces the error of NRPD from two telecom photons
interference, but it reduces the successful entanglement swapping at the same time.

The optimal success probability occurs by using the same excitation efficiency for both cascade and Raman configurations. \ For PNRD, the fidelity is higher
than NRPD, and the heralding probability is the same independent of the degree of frequency space entanglement. \ The success probabilities for both types of
detectors are equal. \ The advantage of PNRD shows up in the fidelity of quantum swapping.

\begin{figure}[t]
\begin{center}
\includegraphics[
natheight=6.00400in,
natwidth=7.00in,
height=2.5239in,
width=3.4809in]%
{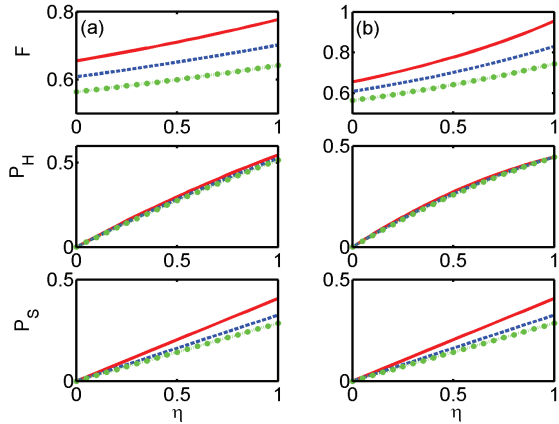}%
\caption{Fidelity $F$, heralding $P_{H}$, and success \ $P_{S}$ probabilities
of entanglement swapping versus telecom detector quantum efficiency $\eta$ for
the case of (a) NRPD and (b) PNRD. \ Solid-red, dashed-blue, and dotted-green
curves correspond to the same parameters used in figure \ref{F_etar}.}%
\label{F_eta}%
\end{center}
\end{figure}

In figure \ref{F_eta}, we show that the measures improve monotonically with
the quantum efficiency ($\eta=\eta_{t}$) of the detector at telecom
wavelength, with $\eta_{r}=0.5$. \ The success probabilities for both types of
detectors are the same and again the advantage of PNRD shows up in the fidelity.%

\subsection{Effective `polarization' maximally entangled (PME) state and quantum teleportation}

\begin{figure}[b]
\begin{center}
\includegraphics[
natheight=6.099600in,
natwidth=7.099800in,
height=2.5086in,
width=3.5421in]%
{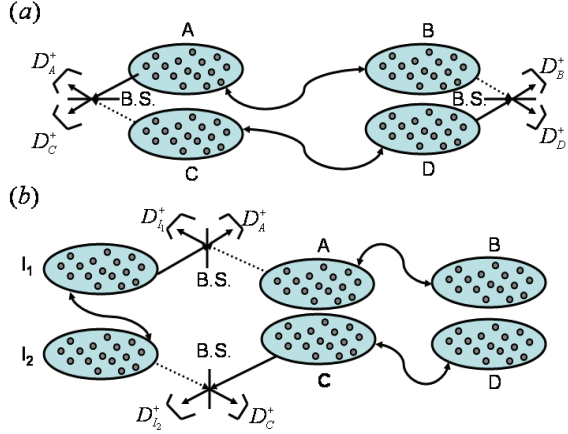}%
\caption{Effective PME projection (a) and quantum teleportation (b) in the DLCZ scheme.
\ Four atomic ensembles\ (A,B,C,D) are used to generate two DLCZ entangled
states at (A,B) and (C,D). \ PME state is projected probabilistically
conditioned on four possible detection events of ($D_{A}^{\dagger}$ or
$D_{C}^{\dagger}$) and ($D_{B}^{\dagger}$ or $D_{D}^{\dagger}$) in (a). \ In
the quantum teleportation protocol (b), another two ensembles (I$_{1},$I$_{2}%
$) are used to prepare a quantum state that is teleported to atomic ensembles
B and D conditioned on four possible detection events of ($\hat{D}_{I_{1}}$ or
$\hat{D}_{A}$) and ($\hat{D}_{I_{2}}$ or $\hat{D}_{C}$).}%
\label{pme_qt}%
\end{center}
\end{figure}

In figure \ref{pme_qt}, we illustrate a scheme for probabilistic and effective PME state preparation and quantum teleportation.\ The term of `polarization' is used as an analogy \cite{dlcz} to the entangled photons in polarization degree of freedom and note that what actually prepared here is the entangled photons in path modes.\ Four ensembles (ABCD) are used to generate two entangled pairs of DLCZ entangled states, and another two ensembles ($I_{1},~I_{2}$) are used to prepare a quantum state to be teleported.%

With the conditional output density matrix from equation (\ref{out}), we proceed to construct the PME state $|\Psi\rangle_{PME}=\frac{1}{\sqrt{2}}(\hat{S}%
_{A}^{\dag}\hat{S}_{D}^{\dag}+\hat{S}_{B}^{\dag}\hat{S}_{C}^{\dag})|0\rangle$ where $(C,D)$ represents another parallel entanglement connection setup,
figure \ref{pme_qt} (a). \ This PME state is useful in entanglement-based communication schemes \cite{dlcz}, and we will here calculate its success
probability. \ The normalized density matrix for the AB\ system is from equation (\ref{out3}) (let $\eta_{t}=\eta$),%

\begin{eqnarray}
\hat{\rho}_{out,n}^{(2),AB} &  =\frac{a}{a+4}|0\rangle\langle0|+\frac{2}%
{a+4}\Big(\hat{S}_{B}^{\dag}|0\rangle\langle0|\hat{S}_{B}+\hat{S}_{A}^{\dag
}|0\rangle\langle0|\hat{S}_{A}\nonumber\\
&  +\lambda_{1}\hat{S}_{B}^{\dag}|0\rangle\langle0|\hat{S}_{A}+\lambda_{1}%
\hat{S}_{A}^{\dag}|0\rangle\langle0|\hat{S}_{B}\Big),
\end{eqnarray}
where the largest Schmidt number ($\lambda_{1}$) of mode overlap is chosen and $a\equiv\eta_{r}(2-\eta)\left(  1+\sum_{j}\lambda_{j}^{2}\right)  $.

A parallel pair of entangled ensembles (C,D) is introduced, and the joint density operator is $\hat{\rho}_{out,n}^{(2),AB}\otimes\hat{\rho}%
_{out,n}^{(2),CD}.$ \ The latter expression is developed mathematically in Appendix B.3.

With projection of the PME state, we have the post measurement success probability [a click from each side; the side of (A or C) and (B or D)],%

\begin{eqnarray}
P_{S,PME}  & =\langle\Psi|\hat{\rho}_{out,n}^{(2),AB}\otimes\hat{\rho}%
_{out,n}^{(2),CD}|\Psi\rangle_{PME},\nonumber\\
& =\frac{4(1+\lambda_{1}^{2})}{[\eta_{r}(2-\eta_{t})(1+\sum_{j}\lambda_{j}%
^{2})+4]^{2}}.
\end{eqnarray}
\ For $\eta_{r}\ll1$, $P_{S,PME}$ reaches the maximum of $1/2$ when a pure source ($\lambda_{1}=1$) is used.\ Compare with the original DLCZ proposal \cite{dlcz} where the success probability is $1/[2(c_0+1)^2]$ with vacuum coefficient $c_0$ in the entanglement generation, we have an equivalent form if a pure source is used, $P_{S,PME}=1/[2(c_0+1)^2]$ where the vacuum coefficient of $\hat{\rho}_{out,n}^{(2),AB}$ can be expressed as $c_0=\eta_r(2-\eta)/2=a/4$.  We may use the PME state to enable the quantum cryptography and Bell inequality measurement by applying phases $\phi_L$ and $\phi_R$ to sides (A, C) and (B, D) in figure \ref{pme_qt}(a) respectively through single-bit operations \cite{dlcz}.  For cascade-emission-based quantum communications, the spectral effect of the cascade emission we implement here reduces the success rate for generation of PME state because of frequency entanglement in the source where $\lambda_1< 1$.

For an arbitrary quantum state transfer to long distance, quantum
teleportation scheme may be used. \ Another two ensembles ($I_{1},I_{2}$) are
introduced \cite{dlcz}, and the quantum state can be described by
$|\Psi\rangle=(d_{0}\hat{S}_{I_{1}}^{\dag}+d_{1}\hat{S}_{I_{2}}^{\dag
})|0\rangle$ with $|d_{0}|^{2}+|d_{1}|^{2}=1$. \ The joint density matrix for
quantum teleportation is%

\begin{equation}
\hat{\rho}_{QT}=(d_{0}\hat{S}_{I_{1}}^{\dag}+d_{1}\hat{S}_{I_{2}}^{\dag
})|0\rangle\langle0|(d_{0}^{\ast}\hat{S}_{I_{1}}+d_{1}^{\ast}\hat{S}_{I_{2}%
})\otimes\hat{\rho}_{out,n}^{(2),AB}\otimes\hat{\rho}_{out,n}^{(2),CD}.
\end{equation}

Atomic ensembles (A,B) in parallel with (C,D) provide a scheme for PME state
preparation. \ Retrieve the quantum state [ensemble ($I_{1},I_{2}$)] into
photons and interfere them at B.S., respectively, with photons from A and C.
We have the teleported quantum state at B and D conditioned on the single
click of ($\hat{D}_{I_{1}}$ or $\hat{D}_{A}$) and ($\hat{D}_{I_{2}}$ or
$\hat{D}_{C}$).

Consider single detection events at $\hat{D}_{I_{1}}$ and $\hat{D}_{I_{2}}$ as
an example. \ With the NRPD measurement operators $\hat{M}_{I_{1},I_{2}}%
\equiv(\hat{I}_{D1}^{\dag}-|0\rangle_{D1}\langle0|)\otimes|0\rangle_{D_{A}%
}\langle0|\otimes(\hat{I}_{D2}^{\dag}-|0\rangle_{D2}\langle0|)\otimes
|0\rangle_{D_{C}}\langle0|$ (we use $D_{1},D_{2}$ for $D_{I_{1}},D_{I_{2}}$),
the density matrix after the measurement becomes%

\begin{eqnarray}\fl
&  \hat{\rho}_{1}\equiv\Tr(\hat{\rho}_{QT,eff}\hat{M}_{I_{1},I_{2}%
})=\nonumber\\\fl
&  \frac{a+2}{2(a+4)^{2}}|0\rangle_{ABCD}\langle0|+\frac{4}{(a+4)^{2}%
}\Big(\frac{|d_{0}|^{2}}{4}\hat{S}_{B}^{\dag}|0\rangle\langle0|\hat{S}%
_{B}+\frac{|d_{1}|^{2}}{4}\hat{S}_{D}^{\dag}|0\rangle\langle0|\hat{S}%
_{D}+\nonumber\\\fl
&  \frac{\lambda_{1}^{2}d_{0}d_{1}^{\ast}}{4}\hat{S}_{B}^{\dag}|0\rangle
\langle0|\hat{S}_{D}+\frac{\lambda_{1}^{2}d_{0}^{\ast}d_{1}}{4}\hat{S}%
_{D}^{\dag}|0\rangle\langle0|\hat{S}_{B}\Big),
\end{eqnarray}
where $\hat{\rho}_{QT,eff}$ is calculated in equation (\ref{QT}), and the trace is
taken over the electromagnetic field degrees of freedom.

For a successful transfer of the quantum state $|\Phi\rangle=(d_{0}\hat{S}%
_{B}^{\dag}+d_{1}\hat{S}_{D}^{\dag})|0\rangle$, the fidelity $F_{1}%
=\langle\Phi|\hat{\rho}_{1}|\Phi\rangle/$Tr$(\hat{\rho}_{1}),$ and the
heralding probability is$~P_{1}=\Tr(\hat{\rho}_{1})$, with the trace
over all atomic degrees of freedom. \ Except for the detection event we
consider here, there are three other detection events including ($D_{A}%
,~D_{C}$), ($D_{I_{1}},~D_{C}$) and ($D_{A},$ $D_{I_{2}}$). \ The teleported
state from the detection events\ ($D_{I_{1}},~D_{C}$) and ($D_{I_{2}},~D_{A}$)
requires a $\pi$ rotation correction\ on the relative phase ($d_{0}\rightarrow
d_{0},$ $d_{1}\rightarrow-d_{1}$).

The fidelity and heralding probabilities conditioned on the other three pairs
of clicks are the same as $F_{1}$ and $P_{1}$ respectively, so the success
probability is
\begin{eqnarray}
P_{S,QT}  & =\sum_{i}^{4}P_{i}F_{i}=4P_{1}F_{1},\nonumber\\
& =\frac{F^{2}}{(1+\lambda_{1})^{2}}[1+(2\lambda_{1}^{2}-2)|d_{0}|^{2}%
|d_{1}|^{2}],
\end{eqnarray}
where $F$ is the fidelity of entanglement swapping for NRPD, equation
(\ref{fidelity}). \ For PNRD, the success probability for quantum
teleportation is unchanged.

\begin{figure}[t]
\begin{center}
\includegraphics[
natheight=6.00400in,
natwidth=7.09700in,
height=2.5462in,
width=3.5105in]%
{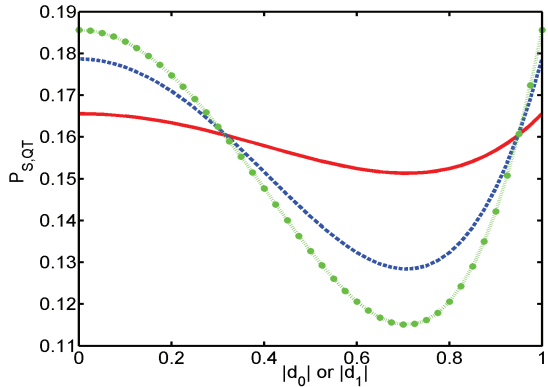}%
\caption{Success probability of quantum teleportation as a function of the
probability amplitude of teleported quantum state with $\eta_{r}=0.5$ and a
perfect detector efficiency $\eta_{t}=1.$ Solid-red, dashed-blue, and
dotted-green curves correspond to the same parameters used in figure
\ref{F_etar}.}%
\label{P_QT}%
\end{center}
\end{figure}

The success probability for quantum teleportation depends on the probability
amplitude of the quantum state and the fidelity $F$ of the entanglement
swapping. \ In figure \ref{P_QT}, for $\eta_{r}=0.5$ and $\eta_{t}=1$, we can
see in the region $|d_{0}|\approx0.3\sim0.9$, higher success probability
requires a less entangled cascade emission source. \ Outside this region, it
prefers a more entangled source. \ When a pure source is used ($\lambda_{1}%
=1$) and let $\eta_{r}\ll1,$ $\eta_{t}=1$, we can achieve the maximum of the
success probability $P_{S,QT}=\frac{1}{4}$ when $F=1$, which is also achieved
in the traditional DLCZ scheme with perfect quantum efficiencies
\cite{shapiro}.%

\section{Discussions and conclusions}

We have described probabilistic protocols for the DLCZ scheme implementing the cascade emission source.\ We characterize the spectral properties of the
cascade emission by Schmidt mode analysis and investigate the fidelity and success probability of the protocols using photon resolving and non-resolving
photon detectors. \ The success probability is independent of the detector type, but photon number resolving \ detection improves the fidelity. \ 

The performance of the protocol also depends on the ratio of efficiencies in generating the cascade and Raman photons. \ The success probability is
optimized for equal efficiencies while the fidelity is higher when the ratio is smaller than one for non-resolving photon detectors.

The frequency space entanglement of telecom photons produced in cascade emission deteriorates the performance of DLCZ protocols. \ The harmful effect
can be diminished by using shorter pump pulses to generate the cascade emission. \ A state dependent success probability of quantum teleportation was
calculated, and in some cases a more highly frequency entangled cascade emission source teleports more successfully. \ An improved performance could
be achieved if the error source (vacuum part) were removed. \ This could be done by entanglement purification \cite{QI2} at the stage of entanglement
swapping and then using the purified source to teleport the quantum state.

The quantum efficiency of detectors have improved to above $60$\% in infra-red wavelength for avalanche photodiodes (APDs) and a maximum of $95$\% in telecom wavelength for superconducting devices at very low temperature (100 mK) \cite{gisin}. \ We expect our optimal performance in the modified DLCZ scheme can be  achieved as shown in figure (\ref{F_eta}) where a fidelity $F\approx 0.9$.\  Our cascade-emission-based quantum communication scheme utilizes a telecom wavelength photon that has minimal loss $0.2$ dB/km through fiber transmission.  Compare with $2$ dB/km loss for infra-red bandwidth, telecom photon has an attenuation length ten times longer which is about $22$ km.  In terms of the rate of direct single photon transmssion over continental distances (several hundreds kilometers), it is overwhelmingly desirable to use telecom over infra-red bandwidth \cite{gisin}.

We note that an alternative method to generate telecom photons in atomic ensembles is frequency down conversion \cite{radaev}. \ Two cold and non-degenerate rubidium gas samples are used to correlate a stored atomic excitation and a telecom photon.\  The stored excitation is correlated with an infra-red photon (idler) in one sample, and the idler is converted to a telecom wavelength photon in the other ensemble. \ Thus a matter-light entanglement is created to serve as a basic element in entanglement connection of DLCZ scheme with an advantageous telecommunication bandwidth.\ Similar to our cascade emission scheme, frequency conversion also requires a phase-matching of four-wave mixing condition in a diamond configuration of atomic levels \cite{telecom, diamond}.\ To implement it into our modified DLCZ scheme, an extra conversion efficiency needs to be taken into account.\ The efficiency has reached a maximum of $0.54$ \cite{radaev} and can be close to one if we use atoms with larger optical depth (opd$>200$) \cite{conversion}.\ Therefore this alternative method serves as well as our cascade emission scheme but demands one more cold atomic ensemble which might cause difficulties when large scale quantum repeater is considered.

\ack
We acknowledge support from NSF, USA and NSC, Taiwan, R. O. C., and thank T. A. B. Kennedy for guidance of this work.

\appendix

\section{Hamiltonian and Schr\"{o}dinger Equation}

In this appendix, we derive the Hamiltonian for the cascade emission
(signal-idler) from a four-level atomic ensemble. \ We use Schr\"{o}dinger's
equation to study the correlated two-photon state from a two-photon laser
excitation.  Consider an ensemble of N four-level atoms interacting with two classical
fields and spontaneously emitted signal and idler photons as shown in figure
\ref{four}. \ These identical atoms distribute randomly with a uniform
density. \ Use dipole approximation of light-matter interactions, $-\vec
{d}\cdot\vec{E}$ where $\vec{E}$ is classical or quantum electric field, and
rotating wave approximation (RWA) \cite{QO:Scully}, the Hamiltonian in interaction picture is
\begin{eqnarray}\fl
&V_I(t)=-\hbar\Delta_1\sum^N_{\mu=1} |1\rangle_\mu\langle 1|-\hbar\Delta_2\sum^N_{\mu=1} |2\rangle_\mu\langle 2|
-\frac{\hbar}{2}\Big\{\Omega_a\hat{P}^\dag_{1\vec{k}_a}+ \Omega_b \hat{P}^\dag_{2\vec{k}_b}+h.c. \Big\}\nonumber\\\fl&-i\hbar\Big\{
\sum_{k_s,\lambda_s}g_{k_s}(\epsilon_{k_s,\lambda_s}\cdot\hat{d}_s^*)\hat{a}_{k_s,\lambda_s}\hat{S}^\dag_{\vec{k}_s}e^{-i(\omega_{ks}-\omega_{23}-\Delta_2)t}\nonumber\\\fl&+\sum_{k_i,\lambda_i}g_{k_i}(\epsilon_{k_i,\lambda_i}\cdot\hat{d}_i^*)\hat{a}_{k_i,\lambda_i}\hat{I}^\dag_{\vec{k}_i}e^{-i(\omega_{ki}-\omega_3)t}
-h.c.\Big\}\label{H},
\end{eqnarray}
where the collective dipole operators, and positive frequency parts of the electric fields are defined as

\begin{eqnarray}
&&\hat{P}^\dag_{1\vec{k}_a}\equiv\sum_{\mu}|1\rangle_\mu\langle 0|
e^{i\vec{k}_a\cdot\vec{r}_\mu},~
\hat{P}^\dag_{2\vec{k}_b}\equiv\sum_{\mu}|2\rangle_\mu\langle 1|e^{i\vec{k}_b\cdot\vec{r}_\mu},\nonumber\\
&&\hat{S}^\dag_{\vec{k}_s}\equiv\sum_{\mu}|2\rangle_\mu\langle 3|
e^{i\vec{k}_s\cdot\vec{r}_\mu},~
\hat{I}^\dag_{\vec{k}_i}\equiv\sum_{\mu}|3\rangle_\mu\langle 0|
e^{i\vec{k}_i\cdot\vec{r}_\mu}\label{op},\\
&&\hat{E}_{s}^{+}(\vec{r}_{1},t_{1})=\sum_{k_{s},\lambda_s}\sqrt{\frac
{\hbar\omega_{s}}{2\epsilon_{0}V}}\hat{a}_{k_{s},\lambda_s}\vec{\epsilon}%
_{k_{s},\lambda_{s}}e^{i\vec{k}_{s}\cdot\vec{r}_{1}-i\omega_{s}t_{1}},\nonumber\\
 &&\hat{E}_{i}^{+}(\vec{r}_{2},t_{2})=\sum_{k_{i},\lambda_i}\sqrt{\frac
{\hbar\omega_{i}}{2\epsilon_{0}V}}\hat{a}_{k_{i},\lambda_i}\vec{\epsilon}%
_{k_{i},\lambda_{i}}e^{i\vec{k}_{i}\cdot\vec{r}_{2}-i\omega_{i}t_{2}}.
\end{eqnarray}

The time dependence of optical frequency in driving fields are absorbed by signal and idler fields.  Single photon detuning $\Delta_1=\omega_a-\omega_1$, and
two-photon detuning $\Delta_2=\omega_a+\omega_b-\omega_2$ ,$\omega_{23}=\omega_2-\omega_3$. Rabi frequencies are
$\Omega_a\equiv (1||\hat{d}||0)\mathcal{E}(k_a)/\hbar$, $\Omega_b\equiv (2||\hat{d}||1)\mathcal{E}(k_b)/\hbar$ and coupling coefficients are
$g_{ks}\equiv(3||\hat{d}||2)\mathcal{E}(k_s)/\hbar$, $g_{ki}\equiv(0||\hat{d}||3)\mathcal{E}(k_i)/\hbar$.  The double matrix element of the dipole moment
is independent of the hyperfine structure and $\mathcal{E}(k)=\sqrt{\frac{\hbar kc}{2\epsilon_0 V}}$.  Polarizations of signal and idler fields
$\epsilon_{k_s,\lambda_s}$, $\epsilon_{k_i,\lambda_i}$ and unit direction of dipole operators $\hat{d}_{s}$, $\hat{d}_{i}$.

In the limit of large detuned and weak driving fields which satisfy $\Delta_1\gg\frac{\sqrt{N}|\Omega_a|}{2}$,
we consider only single excitation and ignore spontaneous decay during excitation.  The state function can be described by

\begin{eqnarray}\fl
|\psi(t)\rangle&=\mathcal{E}(t)|0,vac\rangle+\sum^N_{\mu=1} A_\mu(t)|1_\mu,vac\rangle+
\sum^N_{\mu=1} B_\mu(t)|2_\mu,vac\rangle\nonumber\\\fl&+\sum^N_{\mu=1}\sum_{k_s,\lambda_s}C^\mu_s(t)|3_\mu,1_{\vec{k}_s,\lambda_s}\rangle +
\sum_{k_s,\lambda_s,k_i,\lambda_i}D_{s,i}(t)|0,1_{\vec{k}_s,\lambda_s},1_{\vec{k}_i,\lambda_i}\rangle,
\end{eqnarray}
where $s=(k_s,\lambda_s)$, $i=(k_i,\lambda_i)$, $|m_\mu\rangle\equiv|m_\mu\rangle|0\rangle^{\otimes N-1}_{\nu\neq\mu}$, $m=1,2,3$ and
$|vac\rangle$ is the vacuum photon state.  The probability amplitudes coupled from rotating wave
terms in the Hamiltonian are $\mathcal{E}(t),$ $A_{\mu}(t),$ $B_{\mu}(t),$
$C_{s}^{\mu}(t),$ $D_{s,i}(t),$\ which indicate the complete cycle of single
excitation process from the ground state, intermediate, upper excited state,
intermediate excited state with emission of a signal photon, and the ground
state with the signal-idler emission.  Apply Schr\"{o}dinger equation $i\hbar\frac{\partial}{\partial t}|\psi(t)\rangle=V_I(t)|\psi(t)\rangle$
and we have the coupled equations of motion,
\begin{eqnarray}\fl
i\dot{\mathcal{E}}&=-\frac{\Omega_a^*}{2} \sum_\mu e^{-i\vec{k}_a\cdot\vec{r}_\mu}A_\mu\label{eqn1},\\\fl
i\dot{A}_\mu&=-\frac{\Omega_a}{2}e^{i\vec{k}_a\cdot\vec{r}_\mu}\mathcal{E}-\frac{\Omega_b^*}{2} e^{-i\vec{k}_b\cdot\vec{r}_\mu}B_\mu-
\Delta_1 A_\mu\label{eqn2},\\\fl
i\dot{B}_\mu&=-\frac{\Omega_b}{2}e^{i\vec{k}_b\cdot\vec{r}_\mu}A_\mu-\Delta_2 B_\mu\nonumber\\\fl&-i\sum_{k_s,\lambda_s}g_{ks}
(\epsilon_{k_s,\lambda_s}\cdot\hat{d}_s^*)
e^{i\vec{k}_s\cdot\vec{r}_\mu}e^{-i(\omega_{ks}-\omega_{23}-\Delta_2)t}C^\mu_{s},\label{eqn3}\\\fl
\dot{C}^\mu_{s}&=ig_{ks}^*(\epsilon^*_{k_s,\lambda_s}\cdot\hat{d}_s)e^{-i\vec{k}_s\cdot\vec{r}_\mu}e^{i(\omega_{ks}-\omega_{23}-\Delta_2)t}B_\mu\nonumber\\\fl&-
i\sum_{k_i,\lambda_i}g_{ki}(\epsilon_{k_i,\lambda_i}\cdot\hat{d}_i^*) e^{i\vec{k}_i\cdot\vec{r}_\mu}e^{-i(\omega_{ki}-\omega_3)t}D_{s,i},\label{eqn4}\\\fl
i\dot{D}_{s,i}&=ig_{ki}^*(\epsilon^*_{k_i,\lambda_i}\cdot\hat{d}_i)\sum_\mu e^{-i\vec{k}_i\cdot\vec{r}_\mu}e^{i(\omega_{ki}-\omega_3)t}C^\mu_{s}.\label{eqn5}
\end{eqnarray}

In the limit of large detunings,
\begin{eqnarray}
|\Delta_1|,|\Delta_2|\gg\frac{|\Omega_a|}{2},\frac{|\Omega_b|}{2},\frac{\Gamma_2}{2}\nonumber,
\end{eqnarray}
where $\Gamma_2$ is the natural decay rate for the upper excited state.  We can solve the coupled equations of motion by adiabatically eliminating the intermediate and upper excited states.  The adiabatic approximation \cite{QO:Scully} requires the smoothly turned on of the driving pulses, and it is equivalently to solve for the steady state solutions of the above coupled equations in a perturbative manner.  Since we have weak pump fields, the ground state is constant and other probability amplitudes are
\begin{eqnarray}
A_\mu(t)&\approx& -\frac{\Omega_a(t)}{2\Delta_1}e^{i\vec{k}_a\cdot\vec{r}_\mu},\\
B_\mu(t)&\approx& \frac{\Omega_a(t)\Omega_b(t)}{4\Delta_1\Delta_2}e^{i(\vec{k}_a+\vec{k}_b)\cdot\vec{r}_\mu},
\end{eqnarray}
where probability amplitude of first excited state follows the first driving field and the upper excited state follows the
multiplication of two driving fields.

Substitute equation (\ref{eqn5}) into equation (\ref{eqn4}), we have differential equation for probability amplitude $C^\mu_{s}(t)$,
\begin{eqnarray}\fl
\dot{C}^\mu_{s}(t)&=g_s^*(\epsilon^*_{k_s,\lambda_s}\cdot\hat{d}_s)e^{-i\vec{k}_s\cdot\vec{r}_\mu}e^{i(\omega_{ks}-\omega_{23}-\Delta_2)t}B_\mu(t)
\nonumber\\\fl&-\sum_\nu\sum_{ki,\lambda_i}|g_i|^2|\epsilon_{k_i,\lambda_i}\cdot\hat{d}_i^*|^2e^{i\vec{k}_i\cdot(\vec{r}_\mu-\vec{r}_\nu)}
\int_{0}^t dt'e^{i(\omega_{ki}-\omega_3)(t'-t)}C^\nu_{s}(t').
\end{eqnarray}

Define a phased probability amplitude $C_{s,q_{i}}=\sum_{\mu}C_{s}^{\mu}e^{-i\vec{q}_{i}\cdot\vec{r}_{\mu}}$, substitute $C_{s}^{\nu}$ with $\frac{1}{N}\sum_{q_{i}'} C_{s,q_{i}'}e^{i\vec{q}_{i}'\cdot\vec{r}_{\nu}}$, and identify the terms of the summation of exponential factors, $\sum_\mu e^{i(\vec{k}_i-\vec{q}_i)\cdot\vec{r}_\mu}$ or $\sum_\nu e^{-i(\vec{k}_i-\vec{q}_i')\cdot\vec{r}_\nu}$, the coupling from
the modes $\vec{q}_{i}$ and $\vec{q}_{i}^{\prime}$ are significant only when $|\vec{q}_{i}^{\prime}|=|\vec{k}_{i}|=|\vec{q}_{i}|$, so finally we have 
\begin{equation}\fl
\dot{C}_{s,q_{i}}  =g_{s}^{\ast}(\epsilon_{_{s}}^{\ast}\cdot\hat{d}_{s}%
)\sum_{\mu}e^{-i(\vec{k}_{s}+\vec{q}_{i})\cdot\vec{r}_{\mu}}e^{i(\omega
_{ks}-\omega_{23}-\Delta_{2})t}B_{\mu}-\frac{\Gamma_{3}}{2}(N\bar{\mu}+1)C_{s,q_{i}} +i\delta\omega_{i}C_{s,q_{i}}.%
\end{equation}
The collective decay rate is \cite{Lehm, mu}%

\begin{equation}
\frac{\Gamma_{3}}{2}(N\bar{\mu}+1)\equiv\frac{\Gamma_{3}}{2}\frac{3}{8\pi
}\oint d\Omega_{i}[1-(\hat{k}_{i}\cdot\hat{d}_{i})^{2}]\frac{1}{N}\sum
_{\mu,\nu}e^{i(\vec{k}_{i}-\vec{q}_{i})\cdot(\vec{r}_{\mu}-\vec{r}_{\nu})},
\end{equation}
and the collective frequency shift expressed in terms of the continuous integral over a frequency space is \cite{Lehm, kurizki, Scully2, Scully09}%

\begin{eqnarray}\fl
\delta\omega_{i}  & \equiv&\int_{0}^{\infty}d\omega_{i}\frac{\Gamma_{i}}{2\pi
}\Big[\textrm{P.V.}(\omega_{i}-\omega_{3})^{-1}+\textrm{P.V.}(\omega_{i}%
+\omega_{3})^{-1}\Big]N\bar{\mu}(k_{i}),\\\fl
\bar{\mu}(k_{i})& =&\frac{1}{N^2}\sum_{\mu,\nu\neq\mu}e^{i(\vec{k}_{i}-\vec{q}_{i})\cdot
(\vec{r}_{\mu}-\vec{r}_{\nu})},
\end{eqnarray}
which is derived after we renormalize the Lamb shift and consider the non-RWA terms in the original Hamiltonian.  Non-RWA terms contribute to the term proportional to P.V.$(\omega_{i}+\omega_{3})^{-1}$. 

The geometrical constant $\bar{\mu}$ for a cylindrical ensemble (of height $h
$ and radius $a$) is
\begin{equation}\fl
\bar{\mu}(k_{3})=\frac{6(N-1)}{NA^{2}H^{2}}\int_{-1}^{1}\frac{dx(1+x^{2}%
)}{(1-x)^{2}(1-x^{2})}\sin^{2}[\frac{1}{2}H(1-x)]J_{1}^{2}%
[A(1-x^{2})^{1/2}]\label{mu},%
\end{equation}
where $H=k_{3}h$ and $A=k_{3}a$ are dimensionless length scales, and circular
polarizations are considered \cite{mu}. $J_{1}$ is the Bessel function of the
first kind.

\section{Multimode Description of Correlated Two-Photon State}
In this Appendix, we review a general model for quantum detection
efficiency \cite{det} for multimode analysis in various quantum communication scheme.
\ Based on this detection model with the spectral description of correlated
two-photon state, we derive the effective density matrix conditioning on the
detection events of entanglement swapping, polarization maximally entangled
(PME) state projection, and quantum teleportation.

\subsection{Quantum Efficiency of Detector}

\begin{figure}
\begin{center}
\includegraphics[
natheight=3.099600in,
natwidth=5.099800in,
height=3.0447in,
width=4.5167in]%
{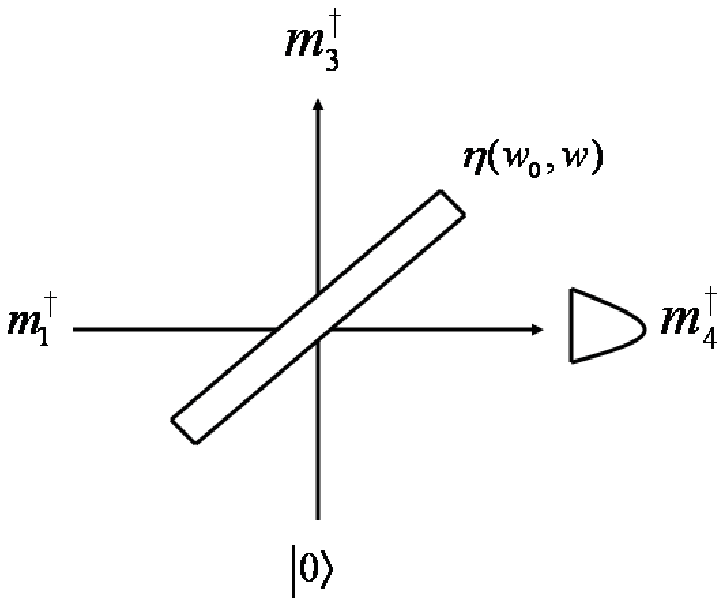}%
\caption{Model of quantum efficiency of detector.}%
\label{detect}%
\end{center}
\end{figure}

To account for quantum efficiency of detector and the affect of its own
spectrum filtering, we introduce an extra beam splitter (B.S.) with a
transmissivity $\eta(\omega,\omega_{0})$ \cite{det} before the detection
event. $\ \eta$ models the quantum efficiency of the detectors in the
microscopic level (response at frequency $\omega_{0}$) and the macroscopic
level (time-integrated detection). \ One example of conditioning on the single
click of the detector, the output density operator becomes%

\begin{eqnarray}
\hat{\rho}_{out}  & =\int_{-\infty}^{\infty}d\omega_{0}\hat{\Pi}_{1}%
\Tr_{ref}\big[\hat{U}_{BS}\hat{\rho}_{in}\hat{U}_{BS}^{\dag}%
\big]\hat{\Pi}_{1}\label{model},\\
\hat{\Pi}_{1}  & \equiv\int_{-\infty}^{\infty}d\omega|\omega\rangle
\langle\omega|,\\
\hat{U}_{BS}  & \equiv\left(
\begin{array}
[c]{cc}%
\sqrt{1-\eta} & \sqrt{\eta}\\
\sqrt{\eta} & -\sqrt{1-\eta}%
\end{array}
\right),
\end{eqnarray}
where $\Tr_{ref}$ is the trace over the reflected modes $m_{3}^{\dag}, $
and the flat spectrum projection operator $\hat{\Pi}_{1}$ (only photon number
is projected and no frequency resolution) is considered in the measurement
process \cite{spectral}. \ In figure \ref{detect}, $m_{1}^{\dag} $ is the
incoming photon operator before the detection, $m_{3}^{\dag}$ is the reflected
mode, and $m_{4}^{\dag}$ is now the detection mode with a modelling of
spectral quantum efficiency and an effective quantum efficiency is defined as%

\begin{equation}
\int_{-\infty}^{\infty}\eta(\omega,\omega_{0})d\omega_{0}=\eta_{eff}(\omega).
\end{equation}

\subsection{Multimode Description of Entanglement Swapping}

From equation (\ref{mode}), we use single mode $\Phi(\omega)$ for Raman photon and
a multimode description $f(\omega_{s},\omega_{i})$ for cascade photons and
rewrite the effective state. \ Note that a symmetric setup is considered so
the mode description is the same for both sides A and B in the scheme of
entanglement swapping.%

\begin{eqnarray}\fl
& |\Psi\rangle_{eff}=\eta_{1}(1-\eta_{2})\times\nonumber\\\fl
& \int f(\omega_{s},\omega_{i})\hat{a}_{s}^{\dag,A}(\omega_{s})\hat{a}%
_{i}^{\dag,A}(\omega_{i})d\omega_{s}d\omega_{i}\int f(\omega_{s}^{\prime
},\omega_{i}^{\prime})\hat{a}_{s}^{\dag,B}(\omega_{s}^{\prime})\hat{a}%
_{i}^{\dag,B}(\omega_{i}^{\prime})d\omega_{s}^{\prime}d\omega_{i}^{\prime
}|0\rangle+\nonumber\\\fl
& \eta_{2}(1-\eta_{1})\int\Phi(\omega)d\omega\hat{a}_{r}^{\dag,A}(\omega
)\hat{S}_{A}^{\dag}\int\Phi(\omega^{\prime})d\omega^{\prime}\hat{a}_{r}%
^{\dag,B}(\omega^{\prime})\hat{S}_{B}^{\dag}|0\rangle+\sqrt{\eta_{1}%
(1-\eta_{1})}\times\nonumber\\\fl
& \sqrt{\eta_{2}(1-\eta_{2})}\int f(\omega_{s},\omega_{i})d\omega_{s}%
d\omega_{i}\times\hat{a}_{s}^{\dag,A}(\omega_{s})\hat{a}_{i}^{\dag,A}%
(\omega_{i})\int\Phi(\omega^{\prime})d\omega^{\prime}\hat{a}_{r}^{\dag
,B}(\omega^{\prime})\hat{S}_{B}^{\dag}|0\rangle+\nonumber\\\fl
& \sqrt{\eta_{1}\eta_{2}(1-\eta_{1})(1-\eta_{2})}\int\Phi(\omega)d\omega
\hat{a}_{r}^{\dag,A}(\omega)\hat{S}_{A}^{\dag}\int f(\omega_{s}^{\prime
},\omega_{i}^{\prime})\hat{a}_{s}^{\dag,B}(\omega_{s}^{\prime})\hat{a}%
_{i}^{\dag,B}(\omega_{i}^{\prime})d\omega_{s}^{\prime}d\omega_{i}^{\prime
}|0\rangle.\nonumber\\\fl&
\end{eqnarray}

With the B.S., we have $\hat{a}_{i}^{\dag,A}=\frac{\hat{m}_{1}^{\dag}+\hat
{m}_{2}^{\dag}}{\sqrt{2}}$, $\hat{a}_{i}^{\dag,B}=\frac{\hat{n}_{1}^{\dag
}+\hat{n}_{2}^{\dag}}{\sqrt{2}}$, $\hat{a}_{r}^{\dag,A}=\frac{\hat{m}%
_{1}^{\dag}-\hat{m}_{2}^{\dag}}{\sqrt{2}}$, $\hat{a}_{r}^{\dag,B}=\frac
{\hat{n}_{1}^{\dag}-\hat{n}_{2}^{\dag}}{\sqrt{2}}$, where $\hat{a}%
_{i}^{\dagger}$ is the creation operator for idler photon and $\hat{a}%
_{r}^{\dagger}$ is for Raman photon. \ The input density operator is
$\hat{\rho}_{in}=|\Psi\rangle_{eff}\langle\Psi|$ and conditioning on the pair
of single click ($\hat{m}_{1,2}^{\dag},\hat{n}_{1,2}^{\dag}$), we are able to
generate maximally entangled singlet or triplet state $|\Psi\rangle
_{DLCZ}=\frac{S_{A}^{\dag}\pm S_{B}^{\dag}}{\sqrt{2}}|0\rangle_{A,B}$. Without
loss of generality, we consider a triplet state along with a pair of clicks
($\hat{m}_{1}^{\dag},\hat{n}_{1}^{\dag}$) and use the model of quantum
efficiency in equation (\ref{model}) with tracing over the detection modes
($\hat{m}_{4}^{\dag},\hat{n}_{4}^{\dag}$). \ Note that $\hat{m}_{1}^{\dag
}=\sqrt{1-\eta}\hat{m}_{3}^{\dag}+\sqrt{\eta}\hat{m}_{4}^{\dag}$ and $\hat
{n}_{1}^{\dag}=\sqrt{1-\eta}\hat{n}_{3}^{\dag}+\sqrt{\eta}\hat{n}_{4}^{\dag}$
as we model the quantum efficiency in the previous Section.%

\begin{eqnarray}\fl
\hat{\rho}_{out}  & =\int_{-\infty}^{\infty}d\omega_{0}\Tr%
_{m4,n4}\big\{\Tr_{m3,n3}\big[\hat{U}_{BS}^{B}\hat{U}_{BS}^{A}\hat{\rho
}_{in}\hat{U}_{BS}^{\dag,A}\hat{U}_{BS}^{\dag,B}\big]\hat{M}_{4,4}\big\},\\\fl
\hat{M}_{4,4}  & \equiv(\hat{I}_{m4}^{\dag}-|0\rangle_{m4}\langle
0|)\otimes|0\rangle_{m2}\langle0|\otimes(\hat{I}_{n4}^{\dag}-|0\rangle
_{n4}\langle0|)\otimes|0\rangle_{n2}\langle0|,
\end{eqnarray}
where the unitary B.S. operator is denoted by both sides (A and B) and NRPD
projection operators are used \cite{shapiro}. \ These operators project the
state with single click of the detected mode without resolving the number of
photons. $\ \hat{I}$ is identity operator. \ The un-normalized output density
operator after tracing out these modes becomes%

\begin{eqnarray}\fl
& \hat{\rho}_{out}=\frac{\eta_{1}^{2}(1-\eta_{2})^{2}}{4}\times\nonumber\\\fl
& \int d\omega_{i}d\omega_{i}^{\prime}\eta_{eff}(\omega_{i})\eta_{eff}%
(\omega_{i}^{\prime})\big[\int f(\omega_{s},\omega_{i})\hat{a}_{s}^{\dag
,A}(\omega_{s})d\omega_{s}\int f(\omega_{s}^{\prime},\omega_{i}^{\prime}%
)\hat{a}_{s}^{\dag,B}(\omega_{s}^{\prime})d\omega_{s}^{\prime}\big]\nonumber\\\fl
& |0\rangle\langle0|\big[\int f^{\ast}(\omega_{s}^{\prime\prime},\omega
_{i})\hat{a}_{s}^{A}(\omega_{s}^{\prime\prime})d\omega_{s}^{\prime\prime}\int
f^{\ast}(\omega_{s}^{\prime\prime\prime},\omega_{i}^{\prime})\hat{a}_{s}%
^{B}(\omega_{s}^{\prime\prime\prime})d\omega_{s}^{\prime\prime\prime
}\big]\nonumber\\\fl
& +\frac{\eta_{1}\eta_{2}(1-\eta_{1})(1-\eta_{2})}{4}\bigg\{\int d\omega
_{i}\eta_{eff}(\omega_{i})\Big[\int f(\omega_{s},\omega_{i})dw_{s}\int
f^{\ast}(\omega_{s}^{\prime},\omega_{i})d\omega_{s}^{\prime}\nonumber\\\fl
& \int|\Phi(\omega)|^{2}\eta_{eff}(\omega)d\omega\Big]\Big(\hat{a}_{s}%
^{\dag,A}(\omega_{s})\hat{S}_{B}^{\dag}|0\rangle\langle0|\hat{S}_{B}\hat
{a}_{s}^{A}(\omega_{s}^{\prime})+\nonumber\\\fl
& \hat{a}_{s}^{\dag,B}(\omega_{s})\hat{S}_{A}^{\dag}|0\rangle\langle0|\hat
{S}_{A}\hat{a}_{s}^{B}(\omega_{s}^{\prime})\Big)+\int\int f(\omega_{s}%
,\omega_{i})d\omega_{s}\Phi^{\ast}(\omega_{i})\eta_{eff}(\omega_{i}%
)d\omega_{i}\times\nonumber\\\fl
& \int\int f^{\ast}(\omega_{s}^{\prime},\omega_{i}^{\prime})d\omega
_{s}^{\prime}\Phi(\omega_{i}^{\prime})\eta_{eff}(\omega_{i}^{\prime}%
)d\omega_{i}^{\prime}\Big(\hat{a}_{s}^{\dag,A}(\omega_{s})\hat{S}_{B}^{\dag
}|0\rangle\langle0|\hat{S}_{A}\hat{a}_{s}^{B}(\omega_{s}^{\prime})+\nonumber\\\fl
& \hat{a}_{s}^{\dag,B}(\omega_{s})\hat{S}_{A}^{\dag}|0\rangle\langle0|\hat
{S}_{B}\hat{a}_{s}^{A}(\omega_{s}^{\prime})\Big)\bigg\}+\hat{\rho}%
_{out}^{\prime},%
\end{eqnarray}
where $\eta_{eff}(\omega)$ is introduced after integration of $\omega_{0}, $
and we denote it as an effective quantum efficiency for idler field
$\omega_{i}$ or Raman photon at frequency $\omega$ (wavelength $780$ nm for D2
line of Rb atom). $\ \hat{\rho}_{out}^{\prime}$ includes the terms that won't
survive after the interference of telecom photons in the middle B.S.
(conditioning on a single click of detector). \ They involve operators like
$\hat{a}_{s}^{\dag,A}\hat{a}_{s}^{\dag,B}|0\rangle\langle0|\hat{a}_{s}^{A}%
\hat{S}_{B}$, $\hat{a}_{s}^{\dag,A}\hat{a}_{s}^{\dag,B}|0\rangle\langle
0|\hat{S}_{A}\hat{S}_{B}$ and $\hat{S}^{\dag,A}\hat{S}^{\dag,B}|0\rangle
\langle0|\hat{S}_{A}\hat{S}_{B}$.

The normalization factor is derived by tracing over the atomic degree of
freedom.
\begin{eqnarray}\fl
& \Tr(\hat{\rho}_{out})\equiv\mathcal{N}=\nonumber\\\fl
& \frac{\eta_{1}^{2}(1-\eta_{2})^{2}}{4}\int d\omega_{s}d\omega_{i}\eta
_{eff}(\omega_{i})|f(\omega_{s},\omega_{i})|^{2}\int d\omega_{s}^{\prime
}d\omega_{i}^{\prime}\eta_{eff}(\omega_{i}^{\prime})|f(\omega_{s}^{\prime
},\omega_{i}^{\prime})|^{2}+\nonumber\\\fl
& \frac{\eta_{1}\eta_{2}(1-\eta_{1})(1-\eta_{2})}{2}\int d\omega_{s}%
d\omega_{i}\eta_{eff}(\omega_{i})|f(\omega_{s},\omega_{i})|^{2}\int|\Phi
|^{2}(\omega)\eta_{eff}(\omega)d\omega+\nonumber\\\fl
& \frac{\eta_{2}^{2}(1-\eta_{1})^{2}}{4}\int|\Phi|^{2}(\omega)\eta
_{eff}(\omega)d\omega\int|\Phi|^{2}(\omega^{\prime})\eta_{eff}(\omega^{\prime
})d\omega^{\prime},\label{normalization}%
\end{eqnarray}
which will be put back when we calculate the heralding and success probabilities.

Next we interfere telecom photons with B.S. that $\hat{a}_{s}^{\dag,A}%
=\frac{\hat{c}_{1}^{\dag}+\hat{c}_{2}^{\dag}}{\sqrt{2}}$, $\hat{a}_{s}%
^{\dag,B}=\frac{\hat{c}_{1}^{\dag}-\hat{c}_{2}^{\dag}}{\sqrt{2}},$ and again a
quantum efficiency $\eta(\omega,\omega_{0})$ for telecom photon is introduced.
\ Use $\hat{c}_{1}^{\dag}=\sqrt{1-\eta}\hat{c}_{3}^{\dag}+\sqrt{\eta}\hat
{c}_{4}^{\dag}$ and trace over the reflected mode $\hat{c}_{3}^{\dag}$
conditioning on the click of $\hat{c}_{4}^{\dag}$ from NRPD \cite{shapiro}. \ The effective
density matrix becomes
\begin{eqnarray}
\hat{\rho}_{out}^{(2)}  & =\int_{-\infty}^{\infty}d\omega_{0}\Tr%
_{c4}\big\{\Tr_{c3}\big[\hat{U}_{BS}^{C}\hat{\rho}_{in}\hat{U}%
_{BS}^{\dag,C}\big]\hat{M}_{4}\big\}\nonumber\\
& \equiv\int_{-\infty}^{\infty}d\omega_{0}\hat{\rho}_{out}^{(2)}(\omega
_{0}),\\
\hat{\rho}_{out}^{(2)}(\omega_{0})  & \equiv\Tr_{c4}\big\{\hat{\rho
}_{in}^{(2)}(\omega_{0})\big\},\\
\hat{M}_{4}  & \equiv(\hat{I}_{c4}^{\dag}-|0\rangle_{c4}\langle0|)\otimes
|0\rangle_{c2}\langle0|,
\end{eqnarray}

\begin{eqnarray}
& \hat{\rho}_{in}^{(2)}(\omega_{0})=\frac{\eta_{1}^{2}(1-\eta_{2})^{2}}%
{16}\int d\omega_{i}d\omega_{i}^{\prime}\eta_{eff}(\omega_{i})\eta
_{eff}(\omega_{i}^{\prime})\bigg\{\nonumber\\
& \int d\omega_{s}(1-\eta(\omega_{s}))f(s,i)f^{\ast}(s,i^{\prime})\int
d\omega_{s}^{\prime}f(s^{\prime},i^{\prime})\sqrt{\eta(\omega_{s}^{\prime}%
)}\hat{c}_{4}^{\dag}(\omega_{s}^{\prime})|0\rangle\langle0|\times\nonumber\\
& \int d\omega_{s}^{\prime\prime}\hat{c}_{4}(\omega_{s}^{\prime\prime}%
)\sqrt{\eta(\omega_{s}^{\prime\prime})}f^{\ast}(s^{\prime\prime},i)+\int
d\omega_{s}(1-\eta(\omega_{s}))f(s,i)f^{\ast}(s,i)\times\nonumber\\
& \int d\omega_{s}^{\prime}f(s^{\prime},i^{\prime})\sqrt{\eta(\omega
_{s}^{\prime})}\hat{c}_{4}^{\dag}(\omega_{s}^{\prime})|0\rangle\langle0|\int
d\omega_{s}^{\prime\prime\prime}\hat{c}_{4}(\omega_{s}^{\prime\prime\prime
})\sqrt{\eta(\omega_{s}^{\prime\prime\prime})}f^{\ast}(s^{\prime\prime\prime
},i^{\prime})+\nonumber\\
& \int d\omega_{s}^{\prime}(1-\eta(\omega_{s}^{\prime}))f(s^{\prime}%
,i^{\prime})f^{\ast}(s^{\prime},i^{\prime})\int d\omega_{s}f(s,i)\sqrt
{\eta(\omega_{s})}\hat{c}_{4}^{\dag}(\omega_{s})|0\rangle\langle
0|\times\nonumber\\
& \int d\omega_{s}^{\prime\prime}\hat{c}_{4}(\omega_{s}^{\prime\prime}%
)\sqrt{\eta(\omega_{s}^{\prime\prime})}f^{\ast}(s^{\prime\prime},i)+\int
d\omega_{s}^{\prime}(1-\eta(\omega_{s}^{\prime}))f(s^{\prime},i^{\prime
})f^{\ast}(s^{\prime},i)\times\nonumber\\
& \int d\omega_{s}f(s,i)\sqrt{\eta(\omega_{s})}\hat{c}_{4}^{\dag}(\omega
_{s})|0\rangle\langle0|\int d\omega_{s}^{\prime\prime\prime}\hat{c}_{4}%
(\omega_{s}^{\prime\prime\prime})\sqrt{\eta(\omega_{s}^{\prime\prime\prime}%
)}f^{\ast}(s^{\prime\prime\prime},i^{\prime})+\nonumber\\
& \int d\omega_{s}^{\prime}\sqrt{\eta(\omega_{s}^{\prime})}f(s^{\prime
},i^{\prime})\int d\omega_{s}\sqrt{\eta(\omega_{s})}f(s,i)\hat{c}_{4}^{\dag
}(\omega_{s})\hat{c}_{4}^{\dag}(\omega_{s}^{\prime})|0\rangle\langle
0|\times\nonumber\\
& \int d\omega_{s}^{\prime\prime}\sqrt{\eta(\omega_{s}^{\prime\prime})}%
f^{\ast}(s^{\prime\prime},i)\int d\omega_{s}^{\prime\prime\prime}\sqrt
{\eta(\omega_{s}^{\prime\prime\prime})}f^{\ast}(s^{\prime\prime\prime
},i^{\prime})\hat{c}_{4}(\omega_{s}^{\prime\prime})\hat{c}_{4}(\omega
_{s}^{\prime\prime\prime})\bigg\}+\nonumber\\
& \frac{\eta_{1}\eta_{2}(1-\eta_{1})(1-\eta_{2})}{8}\bigg\{\int d\omega
_{i}\eta_{eff}(\omega_{i})\int f(s,i)d\omega_{s}\int f^{\ast}(s^{\prime
},i)d\omega_{s}^{\prime}\times\nonumber\\
& \int d\omega|\Phi(\omega)|^{2}\eta_{eff}(\omega)\sqrt{\eta(\omega_{s})}%
\hat{c}_{4}^{\dag}(\omega_{s})\Big(\hat{S}_{B}^{\dag}|0\rangle\langle0|\hat
{S}_{B}+\hat{S}_{A}^{\dag}|0\rangle\langle0|\hat{S}_{A}\Big)\times\nonumber\\
& \hat{c}_{4}(\omega_{s}^{\prime})\sqrt{\eta(\omega_{s}^{\prime})}\int\int
f(s,i)d\omega_{s}\Phi^{\ast}(\omega_{i})\eta_{eff}(\omega_{i})d\omega
_{i}\times\nonumber\\
& \int\int f^{\ast}(s^{\prime},i^{\prime})d\omega_{s}^{\prime}\Phi(\omega
_{i}^{\prime})\eta_{eff}(\omega_{i}^{\prime})d\omega_{i}^{\prime}\sqrt
{\eta(\omega_{s})}\hat{c}_{4}^{\dag}(\omega_{s})\times\nonumber\\
& \Big(\hat{S}_{B}^{\dag}|0\rangle\langle0|\hat{S}_{A}+\hat{S}_{A}^{\dag
}|0\rangle\langle0|\hat{S}_{B}\Big)\hat{c}_{4}(\omega_{s}^{\prime})\sqrt
{\eta(\omega_{s}^{\prime})}\bigg\},
\end{eqnarray}
where a brief notation for spectrum $f(s,i)\equiv f(\omega_{s},\omega_{i})$
and quantum efficiency $\eta(\omega)\equiv\eta(\omega,\omega_{0})$. \ This
quantum efficiency refers to the telecom photon. \ We proceed to trace over
the detected modes and the density matrix can be simplified by interchange of
variables in integration.%

\begin{eqnarray}\fl
& \hat{\rho}_{out}^{(2)}(\omega_{0})=\frac{\eta_{1}^{2}(1-\eta_{2})^{2}}%
{8}\int d\omega_{i}d\omega_{i}^{\prime}\eta_{eff}(\omega_{i})\eta_{eff}%
(\omega_{i}^{\prime})\bigg\{\nonumber\\\fl
& \int d\omega_{s}(1-\eta(\omega_{s},\omega_{0}))f(\omega_{s},\omega
_{i})f^{\ast}(\omega_{s},\omega_{i}^{\prime})\int d\omega_{s}^{\prime}%
f(\omega_{s}^{\prime},\omega_{i}^{\prime})f^{\ast}(\omega_{s}^{\prime}%
,\omega_{i})\eta(\omega_{s}^{\prime},\omega_{0})+\nonumber\\\fl
& \int d\omega_{s}(1-\eta(\omega_{s},\omega_{0}))|f(\omega_{s},\omega
_{i})|^{2}\int d\omega_{s}^{\prime}|f(\omega_{s}^{\prime},\omega_{i}^{\prime
})|^{2}\eta(\omega_{s}^{\prime},\omega_{0})+\nonumber\\\fl
& \frac{1}{2}\int d\omega_{s}^{\prime}\eta(\omega_{s}^{\prime},\omega
_{0})|f(\omega_{s}^{\prime},\omega_{i}^{\prime})|^{2}\int d\omega_{s}%
\eta(\omega_{s},\omega_{0})|f(\omega_{s},\omega_{i})|^{2}+\frac{1}{2}%
\times\nonumber\\\fl
& \int d\omega_{s}^{\prime}\eta(\omega_{s}^{\prime},\omega_{0})f(\omega
_{s}^{\prime},\omega_{i}^{\prime})f^{\ast}(\omega_{s}^{\prime},\omega_{i})\int
d\omega_{s}\eta(\omega_{s},\omega_{0})f(\omega_{s},\omega_{i})f^{\ast}%
(\omega_{s},\omega_{i}^{\prime})\bigg\}|0\rangle\langle0|\nonumber\\\fl
& +\frac{\eta_{1}\eta_{2}(1-\eta_{1})(1-\eta_{2})}{8}\bigg\{\int d\omega
_{i}\eta_{eff}(\omega_{i})\int\eta(\omega_{s},\omega_{0})|f(\omega_{s}%
,\omega_{i})|^{2}d\omega_{s}\times\nonumber\\\fl
& \int d\omega|\Phi(\omega)|^{2}\eta_{eff}(\omega)\Big(\hat{S}_{B}^{\dag
}|0\rangle\langle0|\hat{S}_{B}+\hat{S}_{A}^{\dag}|0\rangle\langle0|\hat{S}%
_{A}\Big)+\nonumber\\\fl
& \int\int\eta(\omega_{s},\omega_{0})f(\omega_{s},\omega_{i})d\omega_{s}%
\Phi^{\ast}(\omega_{i})\eta_{eff}(\omega_{i})d\omega_{i}\int f^{\ast}%
(\omega_{s},\omega_{i}^{\prime})\Phi(\omega_{i}^{\prime})\eta_{eff}(\omega
_{i}^{\prime})d\omega_{i}^{\prime}\times\nonumber\\\fl
& \Big(\hat{S}_{B}^{\dag}|0\rangle\langle0|\hat{S}_{A}+\hat{S}_{A}^{\dag
}|0\rangle\langle0|\hat{S}_{B}\Big)\bigg\},
\end{eqnarray}
where the trace over two photon states requires the commutation relation of
photon operators.%

\begin{eqnarray}
& \Tr[\hat{m}_{4}^{\dag}(\omega_{s})\hat{m}_{4}^{\dag}(\omega
_{s}^{\prime})|0\rangle\langle0|\hat{m}_{4}(\omega_{s}^{\prime\prime})\hat
{m}_{4}(\omega_{s}^{\prime\prime\prime})]\nonumber\\
& =\langle0|\hat{m}_{4}(\omega_{s}^{\prime\prime})[\delta(\omega_{s}%
,\omega_{s}^{\prime\prime\prime})+\hat{m}_{4}^{\dag}(\omega_{s})\hat{m}%
_{4}(\omega_{s}^{\prime\prime\prime})]\hat{m}_{4}^{\dag}(\omega_{s}^{\prime
})|0\rangle,\nonumber\\
& =\delta(\omega_{s},\omega_{s}^{\prime\prime\prime})\delta(\omega_{s}%
^{\prime\prime},\omega_{s}^{\prime})+\delta(\omega_{s},\omega_{s}%
^{\prime\prime})\delta(\omega_{s}^{\prime},\omega_{s}^{\prime\prime\prime}).
\end{eqnarray}

The above is the general formulation for the un-normalized density matrix
conditioning on three clicks of NRPD's. \ We've included spectral quantum
efficiency of the detector either for near-infrared ($\eta_{eff}$) or telecom
wavelength ($\eta_{t}\equiv\int_{-\infty}^{\infty}\eta(\omega,\omega
_{0})d\omega_{0}$)

To proceed, we assume a flat and finite spectrum response ($\eta_{eff}%
(\omega)=\eta_{eff}$, $\eta_{t}(\omega)=\eta_{t}$) with the range $\omega
_{0}\in\lbrack\Omega-\Delta,\Omega+\Delta]$ centered at $\Omega$
(near-infrared or telecom) and $\omega\in\lbrack\omega_{0}-\delta,\omega
_{0}+\delta]$ \cite{det}. \ The widths $2\Delta$ and $2\delta$ are large enough compared
to our source bandwidth so these detection events do not give us any
information of spectrum for our source. \ A perfect efficiency also means no
photon loss during detection. \ Note that the integral involves multiplication
of two telecom photon efficiency $\int_{-\infty}^{\infty}\eta(\omega
,\omega_{0})\eta(\omega^{\prime},\omega_{0})d\omega_{0}=\eta_{t}^{2}(\omega)$
that is valid if the source bandwidth is smaller than detector's.

After the integration of $\omega_{0}$, we have%

\begin{eqnarray}\fl
& \hat{\rho}_{out}^{(2)}=\frac{\eta_{1}^{2}(1-\eta_{2})^{2}}{8}\eta_{eff}%
^{2}\int d\omega_{i}d\omega_{i}^{\prime}\bigg\{(1-\eta_{t})\eta_{t}\int
d\omega_{s}f(\omega_{s},\omega_{i})f^{\ast}(\omega_{s},\omega_{i}^{\prime
})\times\nonumber\\\fl
& \int d\omega_{s}^{\prime}f(\omega_{s}^{\prime},\omega_{i}^{\prime})f^{\ast
}(\omega_{s}^{\prime},\omega_{i})+(1-\eta_{t})\eta_{t}\int d\omega
_{s}|f(\omega_{s},\omega_{i})|^{2}\int d\omega_{s}^{\prime}|f(\omega
_{s}^{\prime},\omega_{i}^{\prime})|^{2}+\nonumber\\\fl
& \frac{\eta_{t}^{2}}{2}\int d\omega_{s}^{\prime}|f(\omega_{s}^{\prime}%
,\omega_{i}^{\prime})|^{2}\int d\omega_{s}|f(\omega_{s},\omega_{i})|^{2}%
+\frac{\eta_{t}^{2}}{2}\int d\omega_{s}^{\prime}f(\omega_{s}^{\prime}%
,\omega_{i}^{\prime})f^{\ast}(\omega_{s}^{\prime},\omega_{i})\times\nonumber\\\fl
& \int d\omega_{s}f(\omega_{s},\omega_{i})f^{\ast}(\omega_{s},\omega
_{i}^{\prime})\bigg\}|0\rangle\langle0|+\frac{\eta_{1}\eta_{2}(1-\eta
_{1})(1-\eta_{2})}{8}\eta_{t}\eta_{eff}^{2}\times\nonumber\\\fl
& \bigg\{\int d\omega_{i}\int|f(\omega_{s},\omega_{i})|^{2}d\omega_{s}\int
d\omega|\Phi(\omega)|^{2}\Big(\hat{S}_{B}^{\dag}|0\rangle\langle0|\hat{S}%
_{B}+\hat{S}_{A}^{\dag}|0\rangle\langle0|\hat{S}_{A}\Big)+\nonumber\\\fl
& \int\int f(\omega_{s},\omega_{i})d\omega_{s}\Phi^{\ast}(\omega_{i}%
)d\omega_{i}\int f^{\ast}(\omega_{s},\omega_{i}^{\prime})\Phi(\omega
_{i}^{\prime})d\omega_{i}^{\prime}\nonumber\\\fl
& \Big(\hat{S}_{B}^{\dag}|0\rangle\langle0|\hat{S}_{A}+\hat{S}_{A}^{\dag
}|0\rangle\langle0|\hat{S}_{B}\Big)\bigg\}.\label{out}%
\end{eqnarray}

\subsection{Density Matrix of PME Projection and Quantum Teleportation}

In Section 4.2, we have the normalized density operator $\hat{\rho}%
_{out,n}^{(2),AB}$\ of the DLCZ entangled state through entanglement swapping.
\ With another pair of DLCZ entangled state, $\hat{\rho}_{out,n}^{(2),CD}$,
the joint density operator for these two pairs constructs the polarization
maximally entangled state (PME) projection and is interpreted as%

\begin{eqnarray}\fl
& \hat{\rho}_{out,n}^{(2),AB}\otimes\hat{\rho}_{out,n}^{(2),CD}=\nonumber\\\fl
& \frac{1}{(a+b)^{2}}\bigg\{a^{2}|0\rangle\langle0|+\frac{ab}{2}%
\Big[|0\rangle_{AB}\langle0|\Big(\hat{S}_{C}^{\dag}|0\rangle\langle0|\hat
{S}_{C}+\hat{S}_{D}^{\dag}|0\rangle\langle0|\hat{S}_{D}\nonumber\\\fl
& +\lambda_{1}\hat{S}_{C}^{\dag}|0\rangle\langle0|\hat{S}_{D}+\lambda_{1}%
\hat{S}_{D}^{\dag}|0\rangle\langle0|\hat{S}_{C}\Big)+|0\rangle_{CD}%
\langle0|\Big(\hat{S}_{B}^{\dag}|0\rangle\langle0|\hat{S}_{B}+\hat{S}%
_{A}^{\dag}|0\rangle\langle0|\hat{S}_{A}\nonumber\\\fl
& +\lambda_{1}\hat{S}_{B}^{\dag}|0\rangle\langle0|\hat{S}_{A}+\lambda_{1}%
\hat{S}_{A}^{\dag}|0\rangle\langle0|\hat{S}_{B}\Big)\Big]+\frac{b^{2}}%
{4}\Big(\hat{S}_{C}^{\dag}|0\rangle\langle0|\hat{S}_{C}+\hat{S}_{D}^{\dag
}|0\rangle\langle0|\hat{S}_{D}\nonumber\\\fl
& +\lambda_{1}\hat{S}_{C}^{\dag}|0\rangle\langle0|\hat{S}_{D}+\lambda_{1}%
\hat{S}_{D}^{\dag}|0\rangle\langle0|\hat{S}_{C}\Big)\otimes\Big(\hat{S}%
_{B}^{\dag}|0\rangle\langle0|\hat{S}_{B}+\hat{S}_{A}^{\dag}|0\rangle
\langle0|\hat{S}_{A}\nonumber\\\fl
& +\lambda_{1}\hat{S}_{B}^{\dag}|0\rangle\langle0|\hat{S}_{A}+\lambda_{1}%
\hat{S}_{A}^{\dag}|0\rangle\langle0|\hat{S}_{B}\Big)\bigg\},
\end{eqnarray}
which is used to calculate the success probability after post measurement [a
click from each side, the side of (A or C) and (B or D)]. \ $a=\eta_{r}%
(2-\eta)\Big(1+\sum_{j}\lambda_{j}^{2}\Big),b=4$, and $\eta_{r}=\eta_{1}%
/\eta_{2}$, $\eta=\eta_{t}$, $\lambda_{j}$ is Schmidt number that is used to
decompose the two-photon source from the cascade transition. \ 

In DLCZ protocol, quantum teleportation uses the similar setup in PME
projection and combines with the desired teleported state, $|\Phi
\rangle=(d_{0}\hat{S}_{I_{1}}^{\dag}+d_{1}\hat{S}_{I_{2}}^{\dag})|0\rangle$,
which is represented by two other atomic ensembles $I_{1}$ and $I_{2}$.
$\ $The requirement of normalization of the state is $|d_{0}|^{2}+|d_{1}%
|^{2}=1$, and the density operator of quantum teleportation is $\hat{\rho
}_{QT}=|\Phi\rangle\langle\Phi|\otimes\hat{\rho}_{out,n}^{(2),AB}\otimes
\hat{\rho}_{out,n}^{(2),CD}$. \ Conditioning on clicks of $\hat{D}_{I_{1}}$
and $\hat{D}_{I_{2}}$, the effective density matrix for quantum teleportation
is (using $\hat{S}_{I_{1}}^{\dag}=(\hat{D}_{I_{1}}+\hat{D}_{A})/\sqrt{2},$
$\hat{S}_{I_{2}}^{\dag}=(\hat{D}_{I_{2}}+\hat{D}_{C})/\sqrt{2}$ for the effect
of beam splitter)%

\begin{eqnarray}\fl
& \hat{\rho}_{QT,eff}=\Big[\frac{|d_{0}|^{2}}{2}(\hat{D}_{I_{1}}^{\dag
}|0\rangle\langle0|\hat{D}_{I_{1}})+\frac{|d_{1}|^{2}}{2}(\hat{D}_{I_{2}%
}^{\dag}|0\rangle\langle0|\hat{D}_{I_{2}})+\frac{d_{0}d_{1}^{\ast}}{2}(\hat
{D}_{I_{1}}^{\dag}|0\rangle\langle0|\hat{D}_{I_{2}})\nonumber\\\fl
& +\frac{d_{0}^{\ast}d_{1}}{2}(\hat{D}_{I_{2}}^{\dag}|0\rangle\langle0|\hat
{D}_{I_{1}})\Big]\otimes\frac{1}{(a+b)^{2}}\bigg\{a^{2}|0\rangle
\langle0|+\frac{ab}{2}\Big[|0\rangle_{AB}\langle0|\nonumber\\\fl
& \Big(\frac{\hat{D}_{I_{2}}^{\dag}|0\rangle\langle0|\hat{D}_{I_{2}}}{2}%
+\hat{S}_{D}^{\dag}|0\rangle\langle0|\hat{S}_{D}+\lambda_{1}\frac{\hat
{D}_{I_{2}}^{\dag}}{\sqrt{2}}|0\rangle\langle0|\hat{S}_{D}+\lambda_{1}\hat
{S}_{D}^{\dag}|0\rangle\langle0|\frac{\hat{D}_{I_{2}}}{\sqrt{2}}%
\Big)\nonumber\\\fl
& +|0\rangle_{CD}\langle0|\Big(\hat{S}_{B}^{\dag}|0\rangle\langle0|\hat{S}%
_{B}+\frac{\hat{D}_{I_{1}}^{\dag}|0\rangle\langle0|\hat{D}_{I_{1}}}{2}%
+\lambda_{1}\hat{S}_{B}^{\dag}|0\rangle\langle0|\frac{\hat{D}_{I_{1}}}%
{\sqrt{2}}+\lambda_{1}\frac{\hat{D}_{I_{2}}^{\dag}}{\sqrt{2}}|0\rangle
\langle0|\hat{S}_{B}\Big)\Big]\nonumber\\\fl
& +\frac{b^{2}}{4}\Big(\frac{\hat{D}_{I_{2}}^{\dag}|0\rangle\langle0|\hat
{D}_{I_{2}}}{2}+\hat{S}_{D}^{\dag}|0\rangle\langle0|\hat{S}_{D}+\lambda
_{1}\frac{\hat{D}_{I_{2}}^{\dag}}{\sqrt{2}}|0\rangle\langle0|\hat{S}%
_{D}+\lambda_{1}\hat{S}_{D}^{\dag}|0\rangle\langle0|\frac{\hat{D}_{I_{2}}%
}{\sqrt{2}}\Big)\otimes\nonumber\\\fl
& \Big(\hat{S}_{B}^{\dag}|0\rangle\langle0|\hat{S}_{B}+\frac{\hat{D}_{I_{1}%
}^{\dag}|0\rangle\langle0|\hat{D}_{I_{1}}}{2}+\lambda_{1}\hat{S}_{B}^{\dag
}|0\rangle\langle0|\frac{\hat{D}_{I_{1}}}{\sqrt{2}}+\lambda_{1}\frac{\hat
{D}_{I_{2}}^{\dag}}{\sqrt{2}}|0\rangle\langle0|\hat{S}_{B}%
\Big)\bigg\},\label{QT}%
\end{eqnarray}
which is used to calculate the success probability for teleported state.


\section*{References}

\begin{thebibliography}{100}

\bibitem {repeater}Briegel H.-J., D\"{u}r W., Cirac J. I. and Zoller P. 1998 {\em Phys. Rev. Lett.} {\bf 81} 5932

\bibitem {Dur}D\"{u}r W., Briegel H.-J., Cirac J. I. and Zoller P. 1999 {\em Phys. Rev. A } {\bf 59} 169

\bibitem{dlcz}Duan L.-M., Lukin M. D., Cirac J. I. and Zoller P. 2001{\em Nature} {\bf 414} 413

\bibitem {qubit}Matsukevich D. N. and Kuzmich A. 2004 {\em Science} {\bf 306} 663 

\bibitem {chou}Chou C. W., Polyakov S. V., Kuzmich A. and Kimble H. J. 2004 {\em Phys. Rev. Lett.} {\bf 92} 213601

\bibitem {store}Chaneli\`{e}re T., Matsukevich D. N., Jenkins S. D., Lan S.-Y., Kennedy T. A. B. and Kuzmich A. 2005 {\em Nature } {\bf 438} 833

\bibitem {pan}Chen S., Chen Y.-A., Strassel T., Yuan Z.-S., Zhao B., Schmiedmayer J. and Pan J.-W. 2006 {\em Phys. Rev. Lett.} {\bf 97} 173004

\bibitem {kimble}Laurat J., de Riedmatten H., Felinto D., Chou C.-W., Schomburg E. W. and Kimble H. J. 2006 {\em Opt. Exp.} {\bf 14} 6912

\bibitem {telecom}Chaneli\`{e}re T., Matsukevich D. N., Jenkins S. D., Lan S.-Y., Zhao R., Kennedy T. A. B. and Kuzmich A. 2006 {\em Phys. Rev. Lett.} {\bf 97} 093604

\bibitem {radaev}Radnaev A. G., Dudin Y. O., Zhao R., Jen H. H., Jenkins S. D., Kuzmich A. and Kennedy T. A. B. 2010 {\em Nature Physics} {\bf 6} 894

\bibitem {conversion}Jen H. H. and Kennedy T. A. B. 2010 {\em Phys. Rev. A} {\bf 82} 023815

\bibitem {clock}Keller T. E. and Rubin M. H. 1997 {\em Phys. Rev. A} {\bf 56} 1534

\bibitem {pulsepump}Grice W. P. and Walmsley I. A. 1997 {\em Phys. Rev. A} {\bf 56} 1627

\bibitem {branning}Branning D., Grice W. P., Erdmann R. and Walmsley I. A. 1999 {\em Phys. Rev. Lett.} {\bf 83} 955

\bibitem {law}Law C. K., Walmsley I. A. and Eberly J. H. 2000 {\em Phys. Rev. Lett.} {\bf 84} 5304

\bibitem {parker}Parker S., Bose S.; and Plenio M. B. 2000 {\em Phys. Rev. A} {\bf 61} 032305

\bibitem {eliminate}Grice W. P., U'Ren A. B. and Walmsley I. A. 2001 {\em Phys. Rev A} {\bf 64} 063815

\bibitem {LOQC}Knill E., Laflamme R. and Milburn G.J. 2001 {\em Nature} {\bf 409} 46

\bibitem {single}U'Ren A B., Silberhorn C., Erdmann R., Banaszek K., Grice W. P., Walmsley I A. and Raymer M. G. 2005 {\em Laser Phys.} {\bf 15} 146

\bibitem {transverse}Law C. K. and Eberly J. H. 2004 {\em Phys. Rev. Lett.} {\bf 92} 127903

\bibitem {micro}Raymer M. G., Noh J., Banaszek K. and Walmsley I. A. 2005 {\em Phys. Rev. A} {\bf 72} 023825

\bibitem {fiber}Garay-Palmett K., McGuinness H. J., Cohen O., Lundeen J. S., Rangel-Rojo R., U'Ren A. B., Raymer M. G., McKinstrie C. J., Radic S. and Walmsley I.
A. 2007 {\em Opt. Express} {\bf 22} 14870

\bibitem{spectral}Humble T. S. and Grice W. P. 2007 {\em Phys. Rev. A} {\bf 75} 022307

\bibitem {QI}Nielsen M. A. and Chuang I. L. 2000 {\em Quantum Computation and Quantum Information} (Cambridge University Press)

\bibitem {cryp}Ekert A. K. 1991 {\em Phys. Rev. Lett.} {\bf 67} 661

\bibitem {QI2}Bouwmeester D., Ekert A. K. and Zeilinger A. 2000 {\em The Physics of Quantum Information: quantum cryptography, quantum teleportation,
quantum computation} (Springer-Verlag Berlin)

\bibitem {Lehm}Lehmberg R. H. 1970 {\em Phys. Rev. A} {\bf 2} 883

\bibitem {Scully09}Scully M. O. 2009 {\em Phys. Rev. Lett.} {\bf 102} 143601

\bibitem {QO:Scully}Scully M. O. and Zubairy M. S. 1997 {\em Quantum Optics} (Cambridge University Press)

\bibitem {QL:Loudon}Loudon R. 2000 {\em The Quantum Theory of Light} (Oxford University Press)

\bibitem {Pan_SPDC} Zhang H., Jin X.-M., Yang J., Dai H.-N., Yang S.-J., Zhao T.-M., Rui J., He Y., Jiang X., Yang F., Pan G.-S., Yuan Z.-S., Deng Y., Chen Z.-B.
, Bao X.-H., Chen S., Zhao B. and Pan J.-W. 2011 {\em Nature Photonics} {\bf 5} 628

\bibitem {0110}van Enk S. J. 2005 {\em Phys. Rev. A} {\bf 72} 064306

\bibitem{shapiro}Razavi M. and Shapiro J. H. 2006 {\em Phys. Rev. A} {\bf 73} 042303

\bibitem{gisin} Sangouard N, Simon C., Riedmatten H. and Gisin N. 2011 {\em Rev. Mod. Phys.} {\bf 83} 33

\bibitem{diamond} Becerra F. E., Willis R. T., Rolston S. L. and Orozco L. A. 2008 {\em Phys. Rev. A} {\bf 78} 013834

\bibitem {mu}Rehler N. E. and Eberly J. H. 1971 {\em Phys. Rev. A} {\bf 3} 1735

\bibitem {kurizki}Mazets I. E. and Kurizki G. 2007 {\em J. Phys. B: At. Mol. Opt. Phys.} {\bf 40} F105

\bibitem {Scully2}Svidzinsky A. A., Chang Jun-Tao and Scully M. O. 2008 {\em Phys. Rev. Lett.} {\bf 100} 160504

\bibitem{det}Rohde P. P. and Ralph T. C. 2006 {\em J. Mod. Opt.} {\bf 53} 1589

\end{thebibliography}
\end{document}